\documentclass[12pt]{article}

\usepackage{amsmath,amsfonts}%,bbold}
\usepackage{latexsym}
\usepackage{amssymb}
\usepackage{epsfig,psfrag,}
\usepackage{stmaryrd}
\usepackage[usenames,dvips]{color}
\usepackage{axodraw}
\usepackage{comment}
\usepackage{cite}

\usepackage{multirow}

\usepackage[margin=3.cm]{geometry}
\makeatletter
\@addtoreset{equation}{section}
\makeatother

\newcommand{\be}{\begin{equation}}
\newcommand{\ee}{\end{equation}}
\newcommand{\bea}{\begin{eqnarray}}
\newcommand{\eea}{\end{eqnarray}}

\newcommand{\h}{\mathfrak h}

\newcommand{\nn}{\nonumber}

\begin{document}
\thispagestyle{empty}

\begin{center}
%\hfill CERN-PH-TH/2009-nnn\\
\hfill CPHT-RR007.0211\\
\hfill UAB-FT-690

\begin{center}

\vspace{.5cm}

{\LARGE\bf Suppressing Electroweak Precision \\ \vspace{.3cm} Observables in 5D Warped Models}

\end{center}

\vspace{1.cm}

{\bf Joan A. Cabrer$^{\,a}$, Gero von Gersdorff$^{\;b}$ 
and Mariano Quir\'os$^{\,c}$}\\

\vspace{1.cm}
${}^a\!\!$ {\em {Institut de F\'isica d'Altes Energies, Universitat Aut{\`o}noma de Barcelona\\
08193 Bellaterra, Barcelona, Spain}}

\vspace{.1cm}

${}^b\!\!$ {\em {Centre de Physique Th\'eorique, \'Ecole Polytechnique and CNRS\\
F-91128 Palaiseau, France}}

\vspace{.1cm}

${}^c\!\!$ {\em {Instituci\'o Catalana de Recerca i Estudis  
Avan\c{c}ats (ICREA) and\\ Institut de F\'isica d'Altes Energies, Universitat Aut{\`o}noma de Barcelona\\
08193 Bellaterra, Barcelona, Spain}}

\end{center}

\vspace{0.8cm}

\centerline{\bf Abstract}
\vspace{2 mm}
\begin{quote}\small
We elaborate on a recently proposed mechanism to suppress large contributions to the electroweak precision observables in five dimensional (5D) warped models, without the need for an extended 5D gauge sector. The main ingredient is a modification of the AdS metric in the vicinity of the infrared (IR) brane corresponding to a strong deviation from conformality in the IR of the 4D holographic dual. We compute the general low energy effective theory of the 5D warped Standard Model, emphasizing additional IR contributions to the wave function renormalization of the light Higgs mode. We also derive expressions for the $S$ and $T$ parameters as a function of a generic 5D metric and zero-mode wave functions. We give an approximate formula for the mass of the radion that works even for strong deviation from the AdS background.
We proceed to work out the details of an explicit model and derive bounds for the first KK masses of the various bulk fields. The radion is the lightest new particle although its mass is already at about 1/3 of the mass of the lightest resonances, the KK states of the gauge bosons.
We examine carefully various issues that can arise for extreme choices of parameters such as the possible reintroduction of the hierarchy problem, the onset of nonperturbative physics due to strong IR curvature or the creation of new hierarchies near the Planck scale. We conclude that a KK scale of 1 TeV is compatible with all these constraints.
\end{quote}

\vfill
 
%Last change: \today

\section{Introduction}
In view of the plethora of forthcoming experimental results expected from the LHC, and given that present precision measurements from the Standard Model (SM) strongly constrain any kind of New Physics, it is urgent to explore all possible models which are consistent with them and which can simultaneously solve (some of) the SM theoretical problems, in particular the hierarchy problem arising from the Higgs sector. Although several mechanisms have been proposed to this end, a particularly elegant one, based and the existence of a warped extra dimension, was suggested by Randall and Sundrum (RS)~\cite{Randall:1999ee}. In the RS model the five-dimensional (5D) space is endowed with an AdS metric and two flat four-dimensional (4D) boundaries, located in the ultraviolet (UV) and infrared (IR) regions, and where the Planck scale is redshifted to the TeV scale by the warp factor. The hierarchy problem is solved in the RS model if the Higgs is localized towards the IR boundary. Using the AdS/CFT correspondence the RS construction is dual to a strongly coupled CFT sector with a UV cutoff where the conformal invariance is spontaneously broken by the IR boundary~\cite{ArkaniHamed:2000ds,Rattazzi:2000hs,PerezVictoria:2001pa}. Moreover fixing the radion vacuum expectation value (VEV) requires introducing a bulk scalar (GW) field with a 5D mass which on the dual side appears as a deformation of the CFT~\cite{Goldberger:1999uk}.

The gauge bosons propagate in the 5D bulk and  their Kaluza-Klein (KK) excitations contribute to the electroweak precision observables. These have to be contrasted with all electroweak precisions tests (EWPT)~\cite{Nakamura:2010zzi}, thus providing lower bounds on the KK-masses. A subset of radiative corrections, the ``oblique corrections", are encoded in a number of observables ($T$, $S$, $W$, $Y$)~\cite{Altarelli:1990zd,Peskin:1991sw,Barbieri:2004qk}  which scale differently with the volume: $T$ grows linearly with the volume while $S$ is volume independent and $Y$ and $W$ are volume suppressed. The $T$ parameter, being the most constraining one, provides bounds on the KK-masses~\cite{Huber:2000fh,Davoudiasl:2009cd} which are outside the LHC reach and create a ``little hierarchy problem", requiring some fine-tuning to stabilize the SM weak masses in the effective theory below the first KK-excitation.  

There are a number of solutions to the large bounds generated by the $T$-parameter within the RS setup. One possibility is to embed the hypercharge in an extended gauge symmetry $SU(2)_R\times U(1)_{B-L}$ in the bulk \cite{Agashe:2003zs}. In this case KK resonances preserve the custodial symmetry $SU(2)_V\subset SU(2)_L\times SU(2)_R$ after electroweak breaking, which protects the $T$ parameter from large tree level corrections. The relevant bounds arise then from the $S$ parameter and turn out to be $\mathcal O(3)$ TeV. Moreover some effort is needed to keep under control volume enhanced corrections to the $Z\bar bb$ coupling~\cite{Agashe:2006at}. Another way of reducing the $T$ parameter in the absence of an extra custodial symmetry is introducing large IR brane kinetic terms~\cite{Davoudiasl:2002ua}. However since IR brane radiative corrections are expected to be small this effect relies on unknown UV physics, which prevents calculability in the low energy effective theory.  

In this paper we will explore another solution to suppress electroweak observables, proposed in Ref.~\cite{Cabrer:2010si}. It consists in replacing the RS metric by a general asymptotically AdS (AAdS) metric in the presence of a bulk scalar field which fixes the brane separation and plays the role of the GW field. AAdS spaces are dual to relevant deformations of the CFT and might give rise to naked singularities in the IR region. {\it Good} singular metrics have been classified and characterized using stringy~\cite{Gubser:2000nd} and pure field theoretical methods~\cite{Cabrer:2009we} and they constitute the background of the so-called soft-wall models~\cite{Cabrer:2009we,AdS/QCD,Falkowski1,Batell2,Delgado:2009xb,MertAybat:2009mk,Atkins:2010cc,vonGersdorff:2010ht}. As we have proven in Ref.~\cite{Cabrer:2010si} if the AAdS metric has a singularity outside the physical interval between the boundaries but nearby the IR brane, and the SM Higgs propagates in the bulk, there is a mechanism which suppresses the electroweak observables. As we will see our results circumvent some general negative analyses in the literature~\cite{Delgado:2007ne,Archer:2010hh} where the SM Higgs was assumed to be localized at the IR brane. On the other hand it was already pointed out that in soft-wall metrics with an extra custodial symmetry in the bulk there is an extra suppression of the $S$ parameter~\cite{Falkowski1} with respect to the RS case.

The outline of this paper is as follows. From Sec.~\ref{sec:5DSM} through Sec.~\ref{radion} we deliver general results which apply to arbitrary 5D metrics and general stabilizing scalar and Higgs backgrounds. In particular in Sec.~\ref{sec:5DSM} we provide the main features and effective theory of the SM propagating in the bulk of the extra dimension. % with an arbitrary 5D metric and Higgs background. 
Similarly general expressions for electroweak precision observables using both the holographic method and integrating out the KK-modes are given in Sec.~\ref{general}. In Sec.~\ref{radion} the gravitational sector in considered for an arbitrary 5D metric and scalar background and an analytical approximation for the light radion mass is provided. From here on we will provide results for a particular AAdS model which deviates from the AdS behavior at the IR region because of the neighborhood of a curvature singularity outside the physical interval. In Sec.~\ref{model} our particular model is introduced and the way the hierarchy is solved is explained in Sec.~\ref{hierarchy}. The main numerical analysis with the bounds on KK-masses based on electroweak observables present data is given in Sec.~\ref{results} and some comments on the soft-wall limit are presented in Sec.~\ref{sw}. Finally the conclusions are drawn in Sec.~\ref{conclusion}. We provide a number of technical details in the appendices. In App.~A details on gauge fluctuations and their equations of motion in the 5D SM are delivered and in App.~B the calculation of the 5D gauge boson propagator is provided. The results from both appendices are widely used in Sec.~\ref{sec:5DSM} and \ref{general}. 

%%%%%%%%%%%%%%%%%%%%%%%%%%%%%%%%%%%%%%%%%%%%%%%%

\section{The 5D Standard Model}
\label{sec:5DSM}

We will now consider the Standard Model (SM) propagating in a 5D space with an arbitrary metric $A(y)$ such that in proper coordinates
\be
ds^2=e^{-2 A(y)}\eta_{\mu\nu}dx^\mu dx^\nu+dy^2\,,
\label{ds2}
\ee
where $\eta_{\mu\nu}=(-1,1,1,1)$ and two flat branes localized at $y=0$ and $y=y_1$, at the edges of a finite interval. We define the 5D $SU(2)_L\times U(1)_Y$ gauge bosons as $W^i_M(x,y)$, $B_M(x,y)$ [or in the weak basis $A_M^\gamma(x,y)$, $Z_M(x,y)$ and $W^{\pm}_M(x,y)$], where $i=1,2,3$ and $M=\mu,5$, and the SM Higgs as
\be
H(x,y)=\frac{1}{\sqrt 2}e^{i \chi(x,y)} \left(\begin{array}{c}0\\h(y)+\xi(x,y)
\end{array}\right)
\ ,
\label{Higgs}
\ee
where the matrix $\chi(x,y)$ contains the three 5D SM fields $\vec\chi(x,y)$, see Eq.~(\ref{chimatrix}). The Higgs background $h(y)$ as well as the metric $A(y)$ will be for the moment arbitrary functions and they will be specified later on.

We will consider the 5D action for the gauge fields, the Higgs field $H$ and other possible scalar fields of the theory, generically denoted as $\phi$:
\begin{eqnarray}
S_5&=&\int d^4x dy\sqrt{-g}\left(-\frac{1}{4} \vec W^{2}_{MN}-\frac{1}{4}B_{MN}^2-|D_M H|^2-\frac12(D_M \phi)^2
-V(H,\phi)
\right)\nonumber\\
&-&\sum_{\alpha}\int d^4x dy \sqrt{-g}\,(-1)^\alpha\,2\,\lambda^\alpha(H,\phi)\delta(y-y_\alpha)
,
\label{5Daction}
\end{eqnarray}
where $V$ is the 5D potential and $\lambda^\alpha$ ($\alpha=0,1$) the 4D brane potentials. From here on we will assume that $V(\phi,H)$ is quadratic in $H$. Electroweak symmetry breaking (EWSB) will be triggered on the IR brane. We thus choose the brane potentials as
\be\lambda^0(\phi_0,H)=M_0 |H|^2
\,,\qquad
-\lambda^1(\phi_1,H)=-M_1 |H|^2+\gamma |H|^4
\,.
\label{boundpot}
\ee
where we denote by $\phi_\alpha$ the VEV of $\phi$ at the brane localized at $y=y_\alpha$.
One can then construct the 4D effective theory out of (\ref{5Daction}) by making the KK-mode expansion
$
A_\mu(x,y)=a_\mu(x)\cdot f_A(y)/\sqrt{y_1}
$
where $A=A^\gamma,Z,W^{\pm}$ and the dot product denotes an expansion in modes. The functions $f_A$ satisfy the equations of motion (EOM)~\footnote{Some details on the spectrum of fluctuations can be found in App.~\ref{fluctuationsgauge}.}
\be
m_{f_A}^2 f_A+(e^{-2A}f^{\prime}_A)'-M_A^2 f_A=0
\ ,
\label{f}
\ee
where the functions $f_A(y)$ are normalized as
$
\int_0^{y_1}f_A^2(y)dy=y_1
$
and satisfy the Neumann boundary conditions (BC)
$
\left. f^{\prime}_A\right|_{y=0,y_1}=0.
$
We have defined the 5D $y$-dependent gauge boson masses as
\be
M_W(y)=\frac{g_5}{2} h(y)e^{-A(y)}
\ ,
\qquad M_Z(y)=\frac{1}{c_W} M_W(y)
\ ,
\qquad M_\gamma(y)
\equiv 0
\ .
\label{masas1}
\ee
where $c_W={g_5}/{\sqrt{g_5^2+g_5'^2}}$, and $g_5$ and $g'_5$ are the 5D $SU(2)_L$ and $U(1)_Y$ couplings respectively. Only the lightest mass eigenvalue will be significantly affected by the breaking so we simplify our notation by defining
\be
m_A=m_{f_A^{0}}\,,%\qquad f_A=f_A^{0}
\ee
for the zero modes and
\be
m_{n}=m_{f_A^{n}}\,,\qquad f^{n}=f^{n}_A\,,
\ee
for the higher modes ($n\geq 1$). In particular masses and wave functions of the $n\geq 1$ KK excitations of the $W$ and $Z$ bosons as well as photon and gluons (almost) coincide.
The masses of light modes ($n=0$) $m_Z$ and $m_W$ have to be matched to the physical values. We will approximately solve the zero mode eigenvalue problem in App.~\ref{fluctuationsgauge}. An approximated expression is given by
\be
m_A^2\approx m_{A,0}^2\equiv\frac{1}{y_1}\int_0^{y_1}dy\, M_A^2(y)\,.
\label{gaugezero}
\ee

On top of the modes from $A_\mu(x,y)$ there will also be pseudoscalar fluctuations $\eta_A$ arising  from the $A_5$ -- $\chi$ sector (see Ref.~\cite{Falkowski1} and App.~\ref{fluctuationsgauge}). For each broken gauge symmetry there is a massive tower of such pseudoscalars.  Their EOM and BC are given by Eqs.~(\ref{eq.eta}) and (\ref{b.c.D}). In the limit of vanishing EWSB they unify with $\xi$ in (\ref{Higgs}) to form complex doublets and hence the splitting for finite breaking is expected to be small (i.~e.~proportional to the mass of the light Higgs).

For the Higgs fluctuations in (\ref{Higgs}) one can write from the action (\ref{5Daction}) an EOM similar to that of gauge bosons (\ref{f}).  The wave functions $\xi(y)$ satisfy the bulk EOM and BC
\be
\xi''(y)-4A'\xi'(y)-\frac{\partial^2 V}{\partial h^2}\xi(y)+m_H^2 e^{2A}\xi(y)=0\,,\qquad 
\frac{\xi'(y_\alpha)}{\xi(y_\alpha)}=\left.\frac{\partial^2 \lambda^\alpha(h)}{\partial h^2}\right|_{y=y_\alpha}
,
\label{Higgsbulk}
\ee
while the background $h(y)$ is determined from 
\be
h''(y)-4A'h'(y)-\frac{\partial V}{\partial h}=0\,,\qquad 
h'(y_\alpha)=\left.\frac{\partial\lambda^\alpha}{\partial h}\right|_{y=y_\alpha}.
\label{bckbulk}
\ee
With our choice of boundary potential, Eq.~(\ref{boundpot}), the UV boundary conditions for background and fluctuations are the same, and 
 for a quadratic bulk Higgs potential the Higgs wave function $\xi(y)$ for $m_H=0$ ($n=0$) is thus proportional to $h(y)$. For small Higgs mass this will still be a good approximation to the exact wave function. This means that the 5D VEV will be carried almost entirely by the zero mode. Let us therefore simplify the discussion by considering an effective theory by writing $H(x,y)=\sqrt k\,\mathcal H(x)h(y)/h(y_1)$ and calculate the effective Lagrangian for the mode $\mathcal H(x)$. Here we have introduced the UV scale $k$ to account for the correct dimension of $\mathcal H$. We will later identify $k$ with the AdS curvature near the UV brane.
 With our choice of boundary potentials, Eq.~(\ref{boundpot}), one finds
\be
\mathcal L_{\rm eff}=- Z e^{-2A_1}\,|D_\mu \mathcal  H|^2-
e^{-4A_1}\left[\left(\frac{h'(y_1)}{h(y_1)}-M_1\right)k|\mathcal H|^2+\gamma k^2|\mathcal H|^4\right]\,.
\label{Leff}
\ee

Several things can be learned from the effective Lagrangian Eq.~(\ref{Leff}). 
The warp factors have the same effect as in the usual RS compactification with a Higgs localized on the IR brane: they red-shift  all mass scales in the IR. The quantity $Z$ is given by
\be
Z=
k\int_0^{y_1}dy\frac{h^2(y)}{h^2(y_1)}e^{-2A(y)+2A(y_1)}\,,
\label{Z}
\ee
which is an additional wave function renormalization depending on both the 
gravitational and Higgs backgrounds.
In Sec.~\ref{general} we will see that a sizable $Z$ can reduce the electroweak precision observables. In particular the $T$ parameter will be suppressed by two powers of $Z$, the $S$ parameter by just one, while the $Y$ and $W$ are unaffected by $Z$. Minimizing the potential in $\mathcal L_{\rm eff}$ one finds the condition
\be
|\langle\mathcal H\rangle|^2=\frac{1}{2\gamma k}\left(M_1-\frac{h'(y_1)}{h(y_1)}\right)\,,
\ee
and hence the physical Higgs mass in the EWSB minimum is
\be
m_H^2
=2 (kZ)^{-1}\left(M_1-\frac{h'(y_1)}{h(y_1)}\right)\rho^2
\ .
\label{mH}
\ee
Here the UV and IR scales, $k$ and $\rho$, are related by~\footnote{This is an obvious generalization of the RS model where $\rho=k e^{-ky_1}$.} 
\be
\rho\equiv k e^{-A(y_1)}\,.
\label{rho}
\ee

Let us now make a few comments on the amount of fine-tuning required to have light modes in the Higgs and  gauge boson sectors. In fact we can write the effective SM Lagrangian as
\be
\mathcal L_{SM}=-\left|\mathcal D_\mu H_{SM}\right|^2+\mu^2|H_{SM}|^2-\lambda |H_{SM}|^4 \,,
\ee
where the SM Higgs field $H_{SM}(x)$ and the SM parameters $\mu^2$ and $\lambda$ are related to 5D quantities by 
\begin{eqnarray}
H_{SM}(x)&=& \sqrt{Z}e^{-A(y_1)} \mathcal H(x)\,,\\
\mu^2&=&(kZ)^{-1}\left(M_1-\frac{h'(y_1)}{h(y_1)}  \right)\; \rho^2\label{muSM}\,,\\
\lambda&=&\frac{\gamma k^2}{Z^2}\label{gammaSM}\,,
\end{eqnarray}
and from where the expressions for the Higgs mass (\ref{mH}) and the gauge boson masses (\ref{gaugezero}) easily follow from the usual SM relations. The required amount of fine-tuning at the tree-level in the 5D parameters is summarized in Eq.~(\ref{muSM}), where we see that depending on the value of $\rho$ and the pre-factor $1/Z$ we have to eventually fine-tune the boundary mass $M_1$ in the factor $M_1-h'(y_1)/h(y_1)$ to obtain $\mu\sim 100$ GeV. As we will see in subsequent sections (see also Ref.~\cite{Cabrer:2010si}), a sizable $Z$ suppresses the $T$ ($S$) parameter by two (one) powers of $Z$, which leads to a corresponding reduction in the value of $\rho$ from EWPT. Moreover we see here that the parameter $\mu$, or equivalently the Higgs mass, is further reduced with respect to $\rho$ by a factor $1/\sqrt{Z}$ which in turn reduces the required amount of fine-tuning in $M_1-h'(y_1)/h(y_1)$. For instance the condition of no fine-tuning would imply that the (dimensionless) parameter
\be
M_1/k-\frac{h'(y_1)}{kh(y_1)}=\frac{m_H^2}{\rho^2}\; \frac{Z}{2}\,,
\label{higgsnoft}
\ee
be of  $\mathcal O(1)$ while a smaller value would imply some amount of fine-tuning, e.~g.~a value of 0.1 (0.01) would amount to a 10\% (1\%) fine-tuning and so on. For a light Higgs this condition imposes strong constraints in our parameter space as we will see in the particular model of Sec.~\ref{model}.

Let us now re-derive Eq.~(\ref{mH}) in a slightly different and more rigorous way.
The relation $\xi(y)=h(y)$  (exact up to normalization for $m_H=0$) 
can be corrected  to $\mathcal O(m_H^2)$. Making an expansion of Eq.~(\ref{Higgsbulk}) yields the corresponding properly normalized wave function~\footnote{We make an expansion in modes $\xi(x,y)= H(x)\cdot \xi(y)/\sqrt{y_1}$ which leads to canonically normalized 4D fields $H(x)$ if we demand $\int  e^{-2A}\,\xi_n^2=y_1$.} 
\be
\xi(y)=\sqrt{\frac{ky_1}{Z}}\frac{h(y)}{h(y_1)}e^{A(y_1)}\left[1-m_H^2\left(\int_0^y e^{2A}\frac{\Omega}{\Omega'}
+\int_0^{y_1} e^{2A}\frac{\Omega}{\Omega'}(\Omega-1)\right)\right]\,
\ ,
\label{xiH}
\ee
where the function $\Omega$ is defined as
\be
\Omega(y)=\frac{\int_0^y h^2(y')e^{-2A(y')}}{\int_0^{y_1}h^2(y')e^{-2A(y')}} \,.
\label{Omega}
\ee
The true value of $m_H$ (and hence the validity of this expansion) is of course determined by the boundary conditions given in Eq.~(\ref{Higgsbulk}). From Eq.~(\ref{xiH}) it follows that
\begin{equation}
\frac{\xi'(y_1)}{\xi(y_1)}  %= 3 \frac{h'(y_1)}{h(y_1)} - 2 M_1
=\frac{h'(y_1)}{h(y_1)}-k Z\frac{m_H^2}{\rho^2}
\label{BCxi}
,
\end{equation}
while from Eq.~(\ref{Higgsbulk}) and (\ref{bckbulk}) we also have
\be
\frac{h'(y_1)}{h(y_1)}=M_1-\gamma h^2(y_1)\,,\qquad 
\frac{\xi'(y_1)}{\xi(y_1)}=M_1-3\gamma h^2(y_1)=
3\frac{h'(y_1)}{h(y_1)}-2M_1\,.
\label{BChiggs}
\ee
Combining Eqs.~(\ref{BCxi}) and (\ref{BChiggs}) one recovers Eq.~(\ref{mH}) for the light Higgs mass as predicted from the effective theory. 
Moreover notice that the BC are universal for all modes and hence Eq.~(\ref{BCxi}) can be used to express the BC for the whole Higgs KK tower in terms of the Higgs mass ($m_H$) rather than the boundary data ($M_1$,\ $\gamma$).

We are finally interested in the coupling of the light Higgs mode to the $W$ and $Z$ bosons and its KK modes, since this will determine how well perturbative unitarity is maintained. In fact having a light Higgs we expect only small corrections to the SM coupling. Using the definition of the $WW\xi_n$ coupling
\be
h_{WW\xi_n}=\frac{g}{y_1}\int_0^{y_1} dy\,e^{-A(y)}M_A(y)f_0^2(y)\xi_n(y) \,, 
\ee
and the wave function (\ref{xiH}) one can deduce that
\be
h_{WW \xi_0}=h_{WW H}^{SM}\left[1-\mathcal O(m_H^2/m_{KK}^2,m_W^2/m_{KK}^2)\right]
,
\ee
so a light Higgs unitarizes the theory in a similar way to the SM Higgs.

%%%%%%%%%%%%%%%%%%%%%%%%%%%%%%%%%%%%%%%%%%%%%%%

\section{General Expressions for Electroweak Precision Observables}
\label{general}

We would now like to present closed expressions for precision observables in arbitrary backgrounds and for arbitrary Higgs profiles. Similar work has been performed in~\cite{Delgado:2007ne,Falkowski1,Falkowski:2009uy,Round:2010kj}. As it is well known~\cite{Kennedy:1988sn} the deviations to electroweak precision measurements will be encoded in the momentum dependence of the propagators of the electroweak gauge bosons. 
For simplicity we will assume for the moment that all the fermions are localized on the UV brane.
The gauge fields that couple to the fermions are thus the brane values of the 5D gauge  fields and hence we will need to calculate the inverse brane-to-brane propagators of the latter. The precision observables can be obtained from these quantities in the straightforward manner~\cite{Barbieri:2004qk}. This is known as the holographic method and will be performed in Sec.~\ref{STholo}. Equivalently one can integrate out the KK modes to obtain effective dimension-six operators involving the fermions and the Higgs. This alternative method, particularly useful in more general settings such as models with bulk fermions, will be presented in Sec.~\ref{KK}.

\subsection{Holographic method}
\label{STholo}
In order to compute the brane-to brane propagator, let us define the quantity 
\be
P(y,p^2,m_{A,0}^2)=e^{-2A(y)}\frac{f_A'(p^2,y)}{f_A(p^2,y)}\,,
\ee
where the holographic profile $f_A(p^2,y)$ satisfies the EOM [see Eq.~(\ref{eqfinalf})]
\be
\left(e^{-2A}f'_A(p^2,y)\right)^\prime=(M_A^2+p^2)f_A(p^2,y) \,.
\ee
From this it follows that $P$ satisfies the differential equation and boundary condition
\be
P'+e^{2A}P^2=p^2+m_{A,0}^2\omega(y)
\label{prop1}
\,,\qquad P(y_1,p^2,m_{A,0}^2)=0 \,,
\ee
with the definitions
\be
m_{A,0}^2=\frac{1}{y_1}\int_0^{y_1}M_A^2(y)\,,
\qquad
\omega(y)=\frac{M_A^2(y)}{m_{A,0}^2}\,.
\label{mA0}
\ee
The function $\omega$ is a distribution which is normalized to $y_1$. For IR (or UV) brane localized Higgses, this function becomes a $\delta$ function supported at the respective boundary.
We will solve Eq.~(\ref{prop1}) in a series expansion in powers of $p^2$ and $m_{A,0}^2$ and finally compute the inverse brane-to-brane propagator
\be
\Pi_A(p^2)=\frac{1}{y_1}P(0,p^2,m_{A,0}^2) \,,
\ee
from which the precision observables can be extracted. The expansion should converge well if the precision observables are small since the suppression scale is expected to be the TeV scale and both the momentum and the quantity $m_{A,0}$ are small compared to that scale~\footnote{In fact the true zero mode mass squared will be given by $m^2_{A}=m^2_{A,0}[1+\mathcal O(m^2_{A,0}/m_{\rm KK}^2)]$.}.
Matching Eq.~(\ref{prop1}) order by order one finds (the subindex denotes the order of both $p^2$ and $m_{A,0}^2$)
\bea
P_0'+e^{2A}P^2_0&=&0\,,\nn\\
P_1'+2e^{2A}P_0P_1&=&p^2+m_{A,0}^2\omega\,,\nn\\
P_2'+e^{2A}(P_1^2+2P_0P_2)&=&0\,.
\eea
Enforcing now the boundary condition at each order one easily finds the solution
\bea
P_0&=&0\,,\nn\\
P_1&=&-p^2(y_1-y)-m_{A,0}^2 y_1 (1-\Omega)\,,\nn\\
P_2&=&\int_y^{y_1} e^{2A}\left[p^2(y_1-y')+m_{A,0}^2y_1(1-\Omega)
\right]^2
\eea
where $\Omega=y_1^{-1}\int_0^y\omega$, explicitly given in Eq.~(\ref{Omega}), is monotonically increasing from $\Omega(0)=0$ to $\Omega(y_1)=1$. In the case of an IR brane localized Higgs it is actually a step function and in particular it vanishes identically in the bulk, $\Omega=0$. 
We end up with the simple expression for the inverse brane-to-brane propagator
\be
\Pi_A(p^2)=-p^2-m_{A,0}^2+y_1
\int_0^{y_1} e^{2A}\left[p^2\left(1-\frac{y}{y_1}\right)+m_{A,0}^2(1-\Omega)\right]^2+\dots \,,
\label{prop2}
\ee
where the dots denote terms of higher order in $p^2$ and $m_{A,0}^2$. This is the quantity from which one can compute all electroweak precision observables related to effective operators of up to dimension six.

All the precision observables can be very easily calculated by applying the above $\Pi_A$'s to various gauge bosons. There are three experimental input parameters (usually referred to as $\epsilon_{1,2,3}$~\cite{Altarelli:1990zd}) that are commonly mapped to the three Peskin-Takeuchi ($S$, $T$, $U$) parameters~\cite{Peskin:1991sw}. However in models with a gap between the electroweak and new physics scales the $U$ parameter is expected to be small since it corresponds to a dimension eight operator. On the other hand, there are dimension six operators such as $(\partial_\mu B_{\nu\sigma})^2$ which in some models can have sizable coefficients and contribute to the $\epsilon_i$.
It has thus been suggested to instead consider the set $T$, $S$, $Y$ and $W$  as a more adequate basis for models of new physics~\cite{Barbieri:2004qk}. They are defined as
\bea
\alpha T&=&
  m_W^{-2}\left[c_W^2 \Pi_{Z}(0)-\Pi_{W}(0)\right]\,,\nn\\
\alpha S&=&
4s_W^2c_W^2\left[\Pi'_{Z}(0)-\Pi_{\gamma}'(0)\right]\,,\nn\\
2m_W^{-2}Y&=&
s_W^2\Pi_{Z}''(0)+c_W^2\Pi_{\gamma}''(0)\nn\,,\\
2m_W^{-2}W&=&
c_W^2\Pi_{Z}''(0)+s_W^2\Pi_{\gamma}''(0)\,,
\eea
where $\alpha$ is the electromagnetic gauge coupling defined at the $Z$-pole mass. The 4D gauge couplings are defined as $g^2=g_5^2/y_1$ and $g'^2=g_5'^2/y_1$.
The quantities $Y$ and $W$ are expected to be relevant whenever the $\mathcal O(p^4)$ in $\Pi(p^2)$ terms cannot be neglected. In theories with a Higgs mode $H$ of mass $m_H\ll m_{\rm KK}$ one can relate $T$, $S$, $Y$, and $W$ to the 
 coefficients of the dimension six operators
\be
|H^\dagger D_\mu H|^2\,,\qquad
H^\dagger W_{\mu\nu}HB^{\mu\nu}\,,\qquad
(\partial_\rho B_{\mu\nu})^2\,,\qquad (D_\rho W_{\mu\nu})^2
\label{ops}
\ee
in the effective low energy Lagrangian respectively. 
Using Eq.~(\ref{prop2}) we can readily calculate these quantities 
\bea
\alpha T&=&s_W^2m_Z^2y_1\int e^{2A}(1-\Omega)^2
=s_W^2 m_Z^2\, \frac{I_2}{\rho^2}\, \frac{ky_1}{Z^2}\,\,,\nn\\
\alpha S&=&8s_W^2c_W^2 m_Z^2 \int e^{2A}\left(y_1-y\right)(1-\Omega)
=8 s^2_W c^2_W m_Z^2\, \frac{I_1}{\rho^2} \,\frac{1}{Z} \,,\nn\\
Y&=&\frac{c_W^2m_Z^{2}}{y_1}
\int e^{2A}\left(y_1-y\right)^2
=c_W^2 m_Z^2\, \frac{I_0}{\rho^2}\,\frac{1}{ky_1}\,,\nn\\
W&=&Y\,,
\label{STYW}
\eea
where we have used the identity
\be
1-\Omega(y)=\frac{u(y)}{Z},\quad u(y)=\int_y^{y_1}dy'\,\frac{h^2(y')}{h^2(y_1)}e^{-2A(y')+2 A(y_1)}
\label{identidad}
\ee
and the notation
\be
I_n=k^{3}\int_0^{y_1} (y_1-y)^{2-n}u^n(y)e^{2A(y)-2A(y_1)},\quad n=0,1,2\,.
\label{notacion}
\ee
The dimensionless integrals $I_n$ are expected to be of the same order. In particular one expects $I_n/\rho^2=\mathcal O(1/m^2_{KK})$ as one can derive these expressions from integrating out the KK modes as it is done in the next subsection. 
We see that $T$ is enhanced by a volume factor and thus it is expected to be the leading observable, while $S$ carries no power of the volume and it is thus the next to leading one. On the other hand the $W$ and $Y$ parameters are suppressed with one (two) additional volume factor(s) compared to $S$ ($T$). In this paper we will thus restrict ourselves to the bounds coming from $T$ and $S$ as those are expected to be the dominant ones and will check a posteriori that the contribution to the $W$ and $Y$ observables is negligible.

The dependence on the $Z$ factors can actually be very easily understood: in Sec.~\ref{sec:5DSM} we saw that they appear in the low-energy effective Lagrangian Eq.~(\ref{Leff}). Therefore the powers of $Z$ in Eq.~(\ref{STYW}) arise in front of the operators in Eq.~(\ref{ops}) by canonically normalizing the Higgs field. We close this section with the observation that $T$ and $S$ parameters are expected to become of the same order when 
\be
\frac{ky_1}{Z} \frac{I_1}{I_2}=8c_W^2\approx 6.2\,.
\ee
Given that the volume is usually $ky_1\sim O(30-35)$ only moderate values of $Z$ are required for this. In the RS model with a bulk Higgs one expects $Z<0.5$.

\subsection{Summing over KK modes}
\label{KK}

In this subsection we present very general expressions for current-current operators which result from integrating out KK gauge bosons of the 5D SM. We will then show that these dimension-six operators result in the same expressions for the precision observables obtained in Sec.~\ref{STholo}. In addition the results we derive will be useful starting points for future studies of more general models with fermions propagating in the bulk, in particular models of flavour~\cite{Agashe:2004cp,Davoudiasl:2009cd}.

\begin{figure}[t]
\begin{center}
%\SetScale{.5}
\begin{picture}(65,80)(35,-20)
\SetWidth{1.}
\Vertex(25,25){3}
\Vertex(100,25){3}
\ArrowLine(0,50)(25,25)
\ArrowLine(25,25)(0,0)
\Photon(25,25)(100,25){3}{5}
\ArrowLine(125,50)(100,25)
\ArrowLine(100,25)(125,0)
\put(30,10){$A_M$}
\put(80,10){$A_N$}
\put(-25,20){$g_5 J_i^M$}
\put(120,20){$g_5J_j^N$}
\end{picture}
\vspace{-1cm}
\end{center}
\caption{\it Diagram contributing to the effective Lagrangian.}
\label{diagramas}
\end{figure}
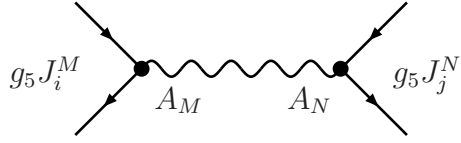

Let us envisage a general gauge interaction of the form
\be
g_5\int d^5x \sqrt{g}\,A_M(x,y) \sum_i  J_{i}^M(x,y)\,,
\ee
where for clarity we have suppressed the sum over the adjoint gauge indices. The index $i$ runs over all the fermions and scalars coupling to $A_M$. Focussing on the zero modes in the currents we can write
\be
J^\mu_i(x,y)=y_1^{-1}e^{2A(y)}\,\omega_i(y) j_i^\mu(x)\,.
\ee
where the functions $\omega_i$ are defined in terms of the zero mode wave functions as
\be
\omega_i(y)\equiv \left\{
\begin{array}{cc}
e^{-2A}\, \xi_{i,0}^2(y)& {\rm scalars}\\
e^{-3A}\, \psi_{i,0}^2(y)& {\rm fermions}
\end{array}\right.,\qquad 
\Omega_i(y)=\frac{1}{y_1}\int_0^y\omega_i\,.
\ee
For later use we also have defined the integral $\Omega_i(y)$. Notice that the functions $\omega_i$ are distributions normalized to $y_1$ which implies that $\Omega_i(y_1)=1$.
We would like to integrate out the KK modes of $A_M$ to obtain the effective Lagrangian of dimension-six operators
\be
\mathcal L_{\rm eff}=\frac{g_5^2}{2} \sum_{i,j}\alpha_{ij} \, j^{\phantom{\mu}}_{i\,\mu}(x) j^\mu_j(x)\,.
\ee
In other words we would like to compute the diagrams in Fig.~\ref{diagramas}.
To this end we need to evaluate 
\be
\alpha^{B_0B_1}_{ij}= \frac{1}{y_1^3}\int_0^{y_1}
 dy\,dy'\,\omega_i(y)\,\omega_j(y')\,G_{B_0B_1}(y,y')
\ee
where $G_{B_0B_1}(y,y')$ is the 5D gauge boson propagator at zero 4D momentum (with any possible zero modes subtracted). The precise definition and explicit expressions are given in App.~\ref{app:propagators}. The subscript $B_i=N,D$ denotes the type of boundary conditions of the gauge field at $y=y_i$ (Neumann or Dirichlet).
Using the results for the propagators, Eqs.~(\ref{GBB}) and (\ref{GNN}), 
one obtains (after performing suitable partial integrations)
\bea
\alpha^{DN}_{ij}&=&\int e^{2A}\,(1-\Omega_i)(1-\Omega_j)\,,\nn\\
\alpha^{ND}_{ij}&=&\int e^{2A}\,\Omega_i\Omega_j\,,\nn\\
\alpha^{DD}_{ij}&=&\int e^{2A}\Omega_i\Omega_j-\frac{\int e^{2A}
\Omega_i\cdot\int e^{2A}\Omega_j}{\int e^{2A}}
\,,\nn\\
\alpha^{NN}_{ij}&=&
\int e^{2A}\,\left(\Omega_i-\frac{y}{y_1}\right)\left(\Omega_j-\frac{y}{y_1}\right)
\,.
\eea
All integrations are between $0$ and $y_1$. These remarkably simple expressions are totally general and valid for arbitrary background metric and matter zero-mode wave function.
It is straightforward to check that in case one or both of the currents are localized on the UV brane ($\Omega_i(y)=1$), $\alpha^{DN}$ and $\alpha^{DD}$ are zero. Similarly if one or both currents reside entirely on the IR brane ($\Omega_i(y)=0$) the amplitudes $\alpha^{ND}$ and $\alpha^{DD}$ vanish. Finally if $\omega_i(y)=1$ (for instance the case of $c=1/2$ for fermions in RS) one finds $\alpha^{NN}=0$. The last result can also be understood from the individual couplings between the current and each gauge KK mode, $\int \omega f_n=\int f_n$. This integral vanishes as a consequence of the orthogonality of the KK wave functions with the (flat) zero mode wave function $f_0=1$. 

Let us apply this formalism to compute the precision observables~\cite{Davoudiasl:2009cd}. Only SM gauge fields propagate in the bulk, and all of them have Neumann-Neumann BC. Imposing a light Higgs it makes sense to keep the latter in the effective theory and compute the operators involving both Higgs and fermion currents. The effective 4D Lagrangian thus reads
\bea
\mathcal L_{\rm eff}=\mathcal L_{\rm SM}
&+&\frac{g_5^2}{2}\left(\widehat\alpha \, j_h^{L}\cdot j^L_{h} +2\widehat\beta\,  j_h^L\cdot j_f^L+\widehat\gamma\,  j_f^L\cdot j_f^L\right)\nn\\
&
+&\frac{g_5'^2}{2}\left( \widehat\alpha\, j_h^{Y}\cdot j^Y_{h} +2\widehat\beta\,  j_h^Y\cdot j_f^Y+\widehat\gamma\,  j_f^Y\cdot j_f^Y\right)
\eea
where $j_h^{L,Y}$ and $j_f^{L,Y}$ are the SM Higgs and fermion currents coupling to $SU(2)_L\times U(1)_Y$ gauge bosons and
\bea
\widehat\alpha&=&\alpha_{hh}^{NN}=\int e^{2A}\,\left(\Omega_h-\frac{y}{y_1}\right)^2
,\nn\\
\widehat\beta&=&\alpha_{hf}^{NN}=\int e^{2A}\,\left(\Omega_h-\frac{y}{y_1}\right)\left(\Omega_f-\frac{y}{y_1}\right)\,,\nn\\
\widehat\gamma&=&\alpha_{ff}^{NN}=\int e^{2A}\,\left(\Omega_f-\frac{y}{y_1}\right)^2
.
\eea

For the time being we have only assumed that all fermions have the same zero mode profile which is however not necessarily localized on the UV brane. The latter case can easily be implemented by setting $\Omega_f(y)=1$. Furthermore since we are working with a light Higgs its wave function will be proportional to the background profile, see Eq.~(\ref{xiH}), and hence $\Omega_h$ coincides with $\Omega$ as defined in Eq.~(\ref{Omega}).
The assumption of ``fermion universality" implies that the effective action only depends on the "complete" fermionic currents $j_f^{L,Y}$ (as opposed to the individual quark and lepton currents) which in turn
allows one to use the gauge boson's EOM to eliminate the fermion currents from the dimension-six operators alltogether~\cite{Davoudiasl:2009cd}. This procedure leaves only the so-called ``oblique corrections"~\cite{Barbieri:2004qk} that in our case take the form~\footnote{The ellipsis denotes operators not relevant to the precision observables.}
\bea
\mathcal L_{\rm eff}=\mathcal L_{\rm SM}
&+&\frac{g_5'^2}{2}(\widehat\alpha-2\widehat\beta+\widehat\gamma)|H^\dagger D_\mu H|^2
+g_5g_5'(\widehat\beta-\widehat\gamma)H^\dagger W_{\mu\nu}HB^{\mu\nu}\nn\\
&+&\frac{\widehat\gamma y_1}{2}(\partial_\rho B_{\mu\nu})^2
+\frac{\widehat\gamma y_1}{2} (D_\rho W_{\mu\nu})^2+\dots
\eea
from which the precision observables 
\bea
\alpha T&=&s_W^2m_Z^2y_1\int e^{2A}(\Omega_f-\Omega)^2\,,\nn\\
\alpha S&=&8s_W^2c_W^2 m_Z^2 y_1\int e^{2A}\left(\Omega_f-\frac{y}{y_1}\right)(\Omega_f-\Omega)\,,\nn\\
Y&=&c_W^2m_Z^{2}\,y_1
\int e^{2A}\left(\Omega_f-\frac{y}{y_1}\right)^2
 \,,\nn\\
W&=&Y\,,
\label{STYW2}
\eea
follow~\footnote{Besides the equality $Y=W$ one can also derive another relation $(\alpha S)^2\leq x (\alpha T) Y$ with $x=64 c_w^2s_w^2\approx 11.2$. This follows from Eq.~(\ref{STYW2}) by use of the Cauchy-Schwarz inequality.}. As expected these expressions reduce to Eqs.~(\ref{STYW}) in the limit of UV-brane localized fermions ($\Omega_f(y)=1$). 
One sees that with $\omega_f=1$ (the case of universal fermion bulk mass $k/2$ in RS) $S$, $W$ and $Y$ are vanishing. On the other hand $T$ is only mildly affected by this choice and does not decrease by much. In order to completely kill the $T$ parameter in RS, one would also need $a=1$ for the Higgs, which corresponds to a mostly elementary Higgs that does not solve the hierarchy problem. On the other hand one could move the fermions more towards the IR and roughly align them with the Higgs profile. In this case the $T$ parameter is suppressed, but $W$ and $Y$ can become dominant, resulting in very high bounds on the KK mass scale. In conclusion, playing with the zero mode wave functions in a pure RS background alone is not enough to reduce the bounds and a modification of the metric background is required.

%%%%%%%%%%%%%%%%%%%%%%%%%%%%%%%%%%%%%%%%%%%%%%%

\section{The Radion}
\label{radion}

Before going on to build a concrete model let us briefly comment on the radion. 
In a slice of pure AdS$_5$ the radion is massless. In the dual theory it is the Goldstone mode of the breaking of scale invariance in the IR. No other scalar modes are present. Adding a stabilizing scalar field corresponds to an explicit breaking of conformal invariance by some relevant operator, the radion becomes a pseudo-Goldstone field and acquires a mass. 
In most cases studied in the literature, the deformation of AdS by the scalar field is small, the radion remains light and its mass can be computed perturbatively~\cite{Csaki:2000zn}. If the deformation of AdS is large (as in the models considered in this paper) one expects a heavy radion.
In this section we will derive an approximation for the radion mass valid in arbitrary backgrounds and carefully analyze under which conditions it is a good one.

The metric and scalar fluctuations can be parametrized as
\begin{gather}
 ds^2 = e^{-2A(y) - 2F(x,y)}  [\eta_{\mu\nu}+ h_{\mu\nu}(x,y)] dx^{\mu} dx^{\nu}  + \left[ 1 + J(x,y) \right]^2dy^2  , 
 \label{eq:metricfluct}
 \\
 \phi(x,y) = \phi(y) + \varphi(x,y) ,
 \label{eq:phifluct}
\end{gather}
where the tensor fluctuations satisfy
\be
h_{\mu\nu}'' - 4A' h_{\mu\nu}' + e^{2A} m^2 h_{\mu\nu} = 0\,,\qquad
h_{\mu\nu}'(y_\alpha)=0\,.
\label{graviton}
\ee
The three scalars $F,\ J,\ \varphi$ are not independent but satisfy the constraints
\begin{gather}
\frac{1}{6}\phi' \varphi = F' - 2 A' F, 
\notag \\
J = 2 F . 
\label{constraint}
\end{gather}
Using these constraints one can decouple the bulk equations and obtain~\cite{Csaki:2000zn}
\be
F'' - 2 A' F' - 4 A'' F  - 2 \frac{\phi''}{\phi'} F' + 4 A' \frac{\phi''}{\phi'} F + m^2 e^{2A} F=0\,,
\label{radionbulk}
\ee
which encompasses both the radion mode as well as all the modes of the stabilizing field.
The boundary condition for the scalar is given in Ref.~\cite{Csaki:2000zn} as
\be
\bigl[-\varphi'+\lambda''_\alpha \varphi+\lambda'_\alpha F\bigr]_{y=y_\alpha}=0\,,
\label{radionBC}
\ee
with the boundary potentials $\lambda_\alpha$ evaluated on the background. Notice that $\lambda_\alpha''(\phi_\alpha)$ is a constant that can be chosen at will without changing the corresponding BC for $\phi$, which only depends on $\lambda_\alpha'(\phi_\alpha)$.
In order to decouple the boundary condition, usually one takes the limit of large $\lambda''_{\alpha}(\phi_\alpha)$ in which case $\varphi$ is frozen at the boundary and Eq.~(\ref{radionBC}) together with the constraint Eq.~(\ref{constraint}) implies $(e^{-2A}F)'|_{y_\alpha}=0$. To be more general we will for the moment not specify the value of $\lambda_\alpha''(\phi_\alpha)$ which remains as a free parameter of the theory. Using the constraint Eq.~(\ref{constraint}), as well as the bulk EOM Eq.~(\ref{radionbulk}), we cast the BC into the form~\footnote{We have also made use of the background EOM $\phi'^2=6A''$ and the BC $\phi'_\alpha=\lambda_\alpha'(\phi_\alpha)$ and hereafter we are using the notation where $X'$ means derivative of $X$ with respect to its dependent variable.}
\be
\left[ m^2 F+\hat M_\alpha(e^{-2A}F)'\right]_{y=y_\alpha}=0\,,
\label{radiongen}
\ee
where we have defined the effective brane mass parameter~\footnote{In terms of the superpotential method used in Sec.~\ref{model} one has $\phi''_\alpha/\phi'_\alpha=W''(\phi_\alpha)$ such that one can express the mass $\hat M_\alpha=\lambda''_\alpha(\phi_\alpha)-W''(\phi_\alpha)$. }
\be
\hat M_\alpha%=\lambda_\alpha''-W''
=\lambda''_\alpha-\frac{\phi_\alpha''}{\phi_\alpha'}\,.
\ee
It is also convenient to recast the bulk EOM Eq.~(\ref{radionbulk}) into the form
\be
\left(e^{2A}(A'')^{-1}[e^{-2A}F]'\right)'+(m^2e^{2A} (A'')^{-1}-2)F=0\,,
\ee
where again the background equations have been used. The new system now only depends on the background metric $A$ and the two free parameters $\hat M_\alpha$. Notice that the quantity $A''(y)$ is a measure of the back reaction and goes to zero in the AdS limit. For $A''=0$ there is thus the expected zero mode $F(y)=e^{2A}$. For small mass we can expand around this mode to obtain the first perturbation
\be
m_0^2=\frac{2\int e^{2A}}{\int e^{4A}(A'')^{-1}+(\hat M_0A_0'')^{-1}-e^{4A_1}(\hat M_1A_1'')^{-1}}\,.
\label{mrad}
\ee
Unless $\hat M_0$ is fine-tuned to zero the second term in the denominator can always be neglected. The third term can however be important and hence the radion mass will depend on it. 
Since we are expanding in the dimensionful parameter $m^2$, the region of validity of this expansion is not so clear. 
To better judge on its convergence one can compute the subleading correction $\delta m_0^2$ to Eq.~(\ref{mrad}). One finds 
\be
\frac{\delta m_0^2}{m_0^2}=-2\int_0^{y_1}e^{2A}\cdot\int_0^{y_1}e^{-2A}A''
\,\chi^2
\label{NLO}
\ee
with the function $\chi(y)$ defined as
\be
\chi(y)=\frac{\int_0^y e^{2A}}{\int_0^{y_1}e^{2A}}-
\frac{\int_0^y e^{4A}(A'')^{-1}}{\int_0^{y_1}e^{4A}(A'')^{-1}-e^{4A_1}(A''_1\hat M_1)^{-1}}\
.
\ee
The important observation is that even if $A''$ is large the correction Eq.~(\ref{NLO}) can be small if the function $\chi(y)$ is small.  For instance consider the limit $\hat M_1\to \infty$. Then $\chi(y)$ is the difference of two functions that monotonically increase from zero to one. This function is thus always smaller than one and, in particular, it vanishes at $y=y_1$ where $A''(y)$ in Eq.~(\ref{NLO}) is expected to be largest. 
For the model to be introduced in Sec.~\ref{model}, $\delta m_0^2/m_0^2$ is always negligible even for radion masses that one would naively consider to be large (i.~e.~of the order of the gauge boson's KK mass). On the other hand the approximation is slightly worse for finite values of $\hat M_1$ and of course breaks down completely when $\hat M_1$ is fine tuned to cancel the first term in the denominator of Eq.~(\ref{mrad}).

Whenever $\delta m_0^2/m_0^2$ is small we expect the couplings of the radion
to be very well approximated by using the leading order wave function 
\be
F(x,y)\simeq e^{2A(y)}r(x)\ .
\ee
This includes but is not limited to the case of very light radion and small back-reaction. We will leave the development of the radion effective theory and its associated phenomenology for future research.

With these remarks we conclude the part of the paper containing the general results. We will now move on to define our particular model.

%%%%%%%%%%%%%%%%%%%%%%%%%%%%%%%%%%%%%%%%%%%%%%%

\section{The Model}
\label{model}

The nontrivial metric of our model will be sourced by a scalar field $\phi$ which 
will also act as a Goldberger-Wise field~\cite{Goldberger:1999uk} stabilizing the distance between the two branes. We will describe the dynamics of the coupled scalar-gravitational system defined by the action
\be
S=\frac{M_5^3}{2}\int d^5x\sqrt{-g}R+S_5\,,
\label{action}
\ee
where $M_5$ is the 5D Planck scale and $S_5$ is given in Eq.~(\ref{5Daction}),
by using the formalism of Ref.~\cite{DeWolfe:1999cp} where first-order gravitational EOM and the bulk potential can be obtained from a superpotential. The 4D (reduced) Planck mass $M_{Pl}=2.4\times 10^{18}$ GeV is related to $M_5$ by
\be
M_{Pl}^2=M_5^3 \int e^{-2A}\,.
\label{relacion}
\ee

We will  introduce on top of the SM Higgs field $H$~\footnote{We will neglect the backreaction of the Higgs field and treat it as an external scalar subject to the gravitational and scalar background. This working hypothesis can be fully justified a posteriori.} the scalar field $\phi$ which will generate a singularity at $y=y_s$, with the superpotential $W(\phi)$ related to the scalar potential by~\footnote{We will set units where $M_5^3=2$ in what follows.}
\be
 V(\phi,h) \equiv \frac{1}{2} 
				\left[ \partial_\phi W(\phi) \right]^2 
			- 
			\frac{1}{3} W(\phi)^2 
	\ +M^2(\phi)|H|^2
	\label{VW}
\ee
Using this ansatz the background EOM can be written as simple first-order differential equations 
\be
A'(y) = \frac{1}{6} W(\phi(y),h(y)) ,\quad
\phi'(y) = \partial_\phi W(\phi,h)
 \label{metricEOM}
\ee
whose solutions will enter the usual second order linear equation for $h(y)$
\be
h''(y)-4A'(y)h'(y)-M^2[\phi(y)]\, h(y)=0\,.
\label{higgsbg}
\ee
In terms of the boundary potentials $\lambda^\alpha(\phi,h)
$ the boundary conditions are
\be
A'(y_\alpha)=\left.\frac{2}{3}\lambda^\alpha(\phi,h)\right|_{y=y_\alpha},\quad
\phi'(y_\alpha)=\left.\frac{\partial\lambda^\alpha}{\partial\phi}\right|_{y=y_\alpha},\quad
h'(y_\alpha)=\left.\frac{\partial\lambda^\alpha}{\partial h}\right|_{y=y_\alpha}\ .
\label{BCs}
\ee

\begin{figure}[tb]
\centering
\begin{psfrags}
\input{L1-psfrag.tex}
\includegraphics[width=0.67\textwidth]{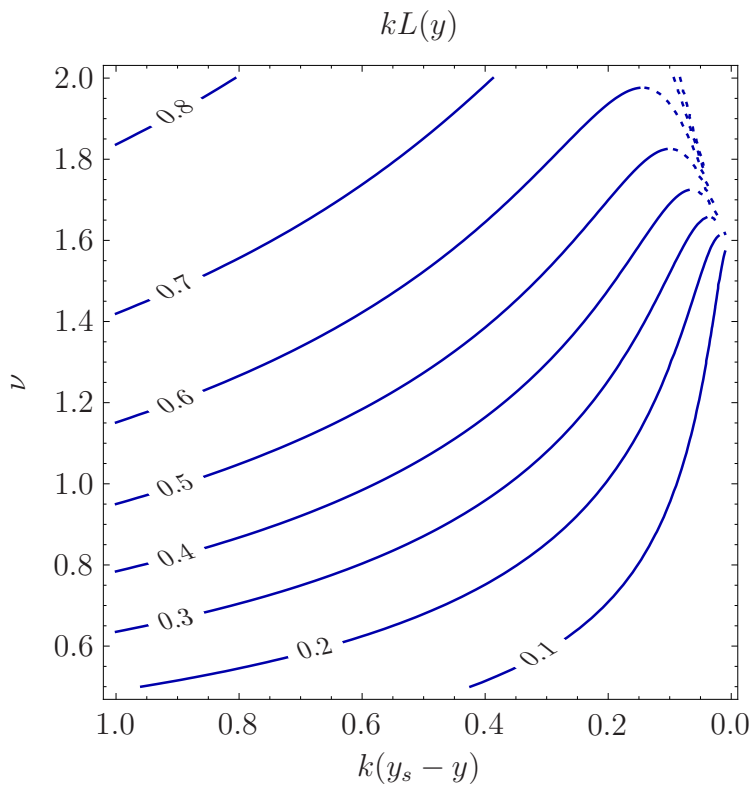}
\end{psfrags}
\caption{\it Contour levels of $k L(y)$ as a function of $\nu$ and $k(y_s - y)$. For $\nu > \sqrt{5/2}$, $L(y)$ presents a minimum close to the singularity. Since we demand $L(y)$ to be a monotonic function we will constrain ourselves outside the region after this minimum is reached (dotted lines).}
\label{fig0}
\end{figure}

The model is thus completely specified by the choice of the two bulk functions $W(\phi)$ and $M^2(\phi)$, as well as by the two brane potentials $\lambda_\alpha(\phi,h)$. We will postulate the following superpotential 
\be
W_\phi(\phi)=6k(1+b e^{\nu\phi/\sqrt{6}})\,,
\label{superp}
\ee
where $\nu$ and $b$ are real arbitrary parameters and $k$ is the inverse curvature radius near the UV brane where the space is almost AdS. This leads to the background configuration~\cite{Cabrer:2009we}
\bea
\phi(y)&=&-\frac{\sqrt{6}}{\nu}\log[\nu^2 b k(y_s-y)]
\label{phi}
\ ,
\\
A(y)&=&ky-\frac{1}{\nu^2}\log\left(1-\frac{y}{y_s}\right)
\ ,
\label{A}
\end{eqnarray}
where we are using the normalization $A(0) =  0$.  Moreover the 5D curvature $R(y)$ and the cuvature radius $L(y)$ along the fifth dimension  which are given by
\be
%%%R(y)=\frac{4}{3}\left(\frac{\partial W}{\partial\phi}\right)^2-\frac{5}{9}W^2
R(y)= 8 A''(y) - 20\left[A'(y)\right]^2
,\quad 
L(y)=\sqrt{\frac{-20}{R}}
\label{curvature}
\ee
 are in our case 
\be
k L(y) = \frac{ \nu^2 k(y_s - y) }{\sqrt{1 - 2 \nu^2/5 + 2 \nu^2k(y_s - y) +  \nu^4k^2(y_s - y)^2}} .
\label{curvature-res}
\ee 
Near the UV brane we have $kL_0\sim 1$~\footnote{This shows that $k$ is approximately the inverse AdS curvature radius.},
and $k L(y)$ remains close to unity in most of the interbrane distance. Near the IR brane $L$ can get small due to the spurious singularity at $y=y_s$.
The quantity $kL_1$ is thus a useful measure of the deviation from pure RS and we will use $kL_1$ instead of $k \Delta$ (the distance between the IR brane and the singularity).
The behavior of $L(y)$ in the IR is shown in Fig.~\ref{fig0}.  $L(y)$ is a monotonically decreasing function when $\nu \le \sqrt{5/2}$ and hence in this region a bound on the curvature implies a lower bound on $\Delta$. When $\nu > \sqrt{5/2}$ the curvature radius possesses a minimum (correspondingly the curvature presents a maximum) and eventually the curvature changes sign before the singularity. Later on we will present our results as a function of $L(y)$ so, for simplicity, we will impose that the IR brane is located before this minimum is reached so that $L(y)$ is a monotonic function. We will see that this constraint only affects mildly the final results, and only in the least interesting region of our model.

We assume that the brane dynamics $\lambda^\alpha_\phi$ fixes the values of the field \mbox{$\phi=(\phi_0,\,\phi_1)$} on the UV and IR branes respectively. The inter-brane distance $y_1$, as well as the location of the singularity at $y_s\equiv y_1+\Delta$ and the warp factor $A(y_1)$, are related to the values of the field $\phi_\alpha$ at the branes by the following simple expressions:
\begin{eqnarray}
ky_1&=&\frac{1}{\nu^2}\left[e^{-\nu \phi_0/\sqrt{6}}-e^{-\nu \phi_1/\sqrt{6}}  \right], \quad k\Delta=\frac{1}{\nu^2}e^{-\nu \phi_1/\sqrt{6}}
\ ,
\nonumber\\
A(y_1)&\simeq&k y_1+\frac{1}{\nu}(\phi_1-\phi_0)/\sqrt{6}
\ ,
\end{eqnarray}
which shows that the required large hierarchy can naturally be 
fixed with values of the fields $\phi_1\gtrsim \phi_0$, $\phi_0<0$ and $\mathcal O(1)$ in absolute value. Moreover the strict soft-wall configuration \cite{Cabrer:2009we} corresponds to the limit $\phi_1\gg 1,\, y_1\to y_s$. 
Also note that due to its exponential dependence on $\phi_1$, $\Delta$ can be small or, in other words, the IR brane can naturally be located very close to the singularity.

As for the Higgs bulk mass term we choose
\be
M^2(\phi)=ak\left[ak-\frac{2}{3}W(\phi)\right]\,.
\label{M2}
\ee
where $a$ is an arbitrary real parameter. As $a$ is constrained by the hierarchy problem we will restrict its range to values $a>2$ as we will see in the next section.
The choice Eq.~(\ref{M2}) ensures that one linearly independent solution to Eq.~(\ref{higgsbg}) is given by a simple exponential. Certainly other choices are possible which will lead to similar results. We will comment on some of them in Sec.~\ref{hierarchy}.

Using the superpotential formalism to define the $\phi$ potential amounts to some fine-tuning among the different coefficients of the bulk potential, unless they are protected by some underlying 5D supergravity~\cite{DeWolfe:1999cp}. The quadratic Higgs term which is generated by (\ref{VW}) can be written as
$k^2[a(a-4)-4 a b e^{\nu\phi/\sqrt{6}} ]|H|^2$
and the coefficients of the two operators $|H|^2$ and $e^{\nu\phi/\sqrt{6}}|H|^2$ can be considered as independent parameters~\footnote{Of course the coefficients of the operators not involving  the Higgs field remain fine-tuned as we are using the superpotential formalism to fix them.}. However since the parameter $b$ can be traded by a global shift in the value of the $\phi$ field, or in particular by a shift in its value at the UV brane $\phi_0$, for simplicity we will fix its value to $b=1$ hereafter.

Having fixed the background we can write the general solution to Eq.~(\ref{higgsbg}) as
\be
h(y)=e^{aky}\left(c_1+c_2\int^y e^{-2(a-2) k y'}\left(1-\frac{y'}{y_s}\right)^{-\frac{4}{\nu^2}}\right)\,.
\ee
The two integration constants $c_i$ are fixed from the boundary conditions Eq.~(\ref{BCs}) derived from the boundary potentials $\lambda_\alpha$ given in Eq.~(\ref{boundpot}). 
To adequately suppress the $T$ parameter we would like to keep the exponential solution (which corresponds to the non-singular solution in the SW limit). In the next section we will see that this imposes some restrictions on the parameter space which will have a simple holographic interpretation.

\begin{figure}[p]
\centering
\begin{minipage}[c]{0.19\textwidth}
~~~~~~~~\textbf{(a)}
\end{minipage}
\begin{minipage}[c]{0.8\textwidth}
 \begin{psfrags}
  \input{Z1-psfrag.tex}
  \includegraphics[width= 0.83\textwidth]{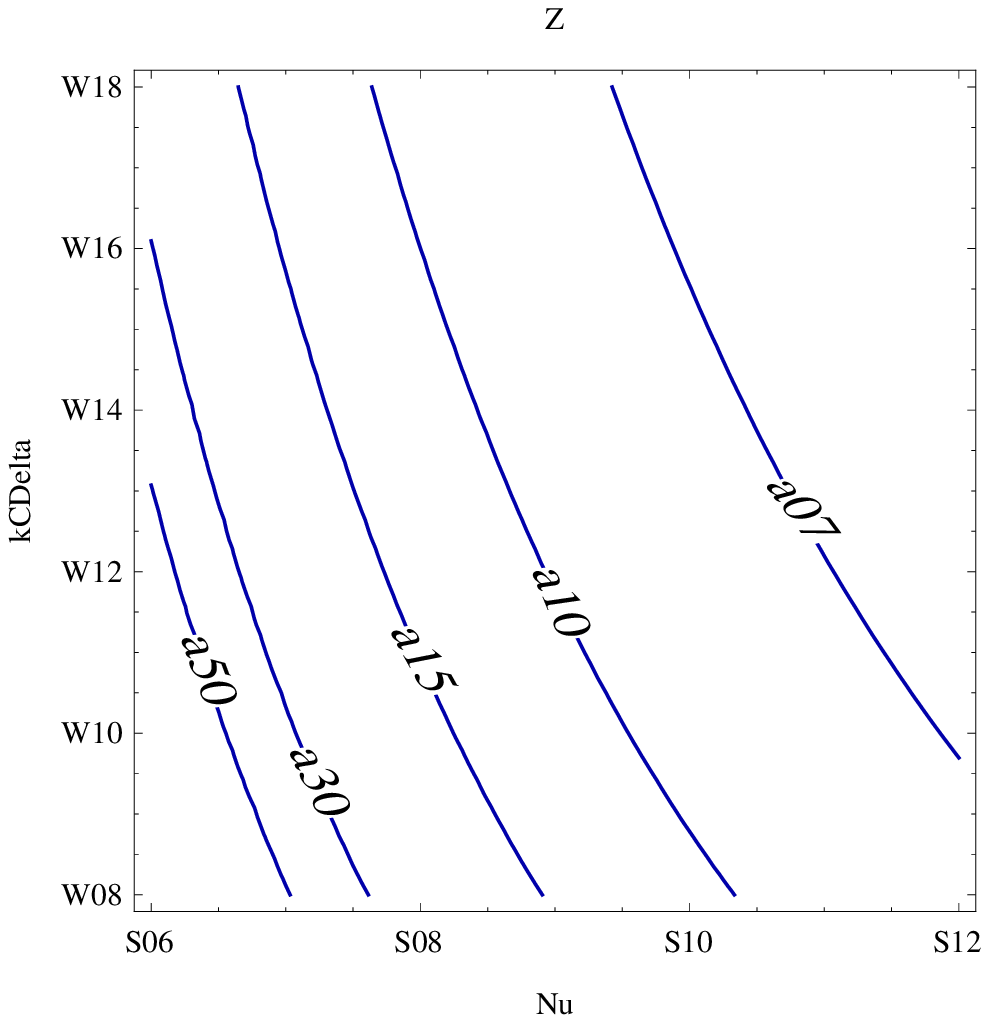}
 \end{psfrags}
\end{minipage}
\\ \vspace{5mm}
\begin{minipage}[c]{0.19\textwidth}
~~~~~~~~\textbf{(b)}
\end{minipage}
\begin{minipage}[c]{0.8\textwidth}
 \begin{psfrags}
  \input{Z2-psfrag.tex}
  \includegraphics[width= 0.83\textwidth]{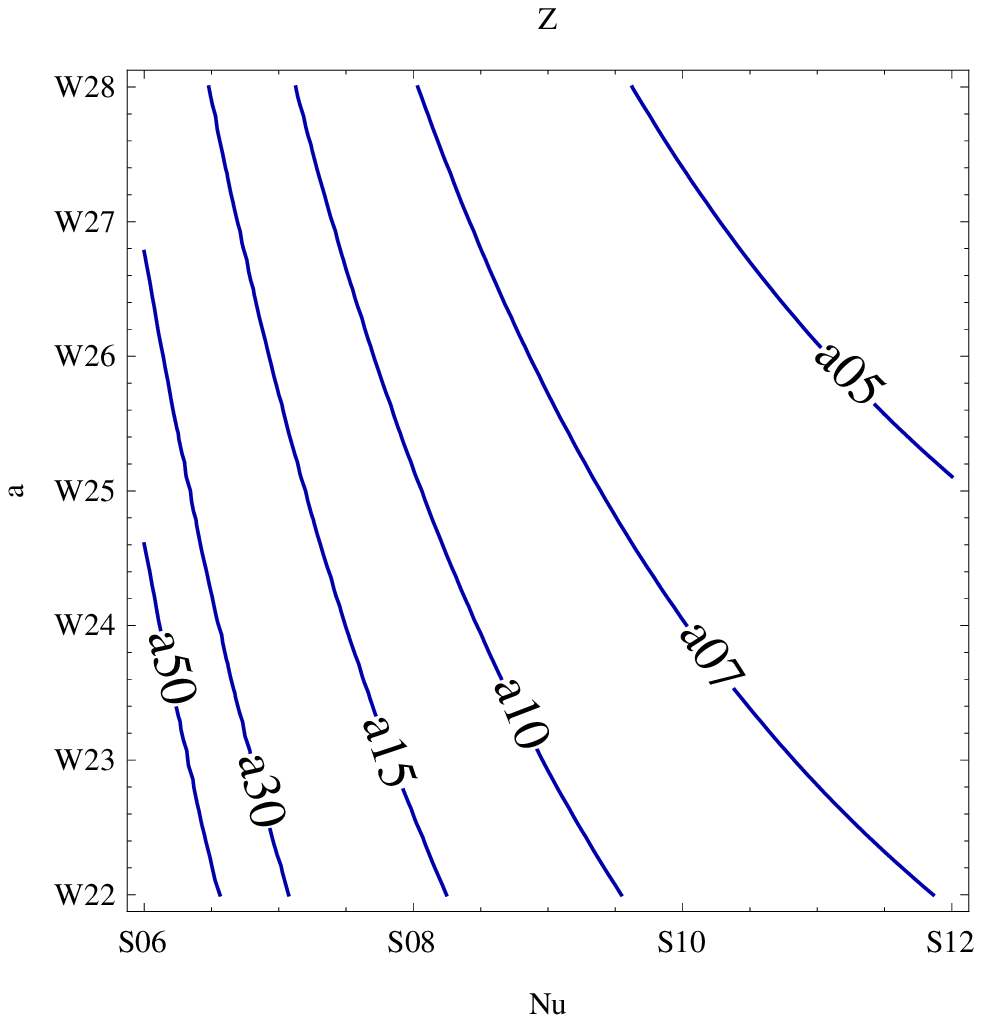}
 \end{psfrags}
\end{minipage}
\caption{\it Contour levels of $Z$ for $A(y_1)=35$: {\bf (a)} as a function of $\nu$ and $k\Delta$ for $a=2.2$; and {\bf (b)} as a function of $\nu$ and $a$ for $k\Delta=1$. }
\label{fig1}
\end{figure}

As it was already stressed a reduction of the $S$ and $T$ parameters can occur provided that the $Z$ factors defined in Eq.~(\ref{Z}) are sizable, and we will present numerical results for the model in Sec.~\ref{results}. In fact a quick glance at Eqs.~(\ref{STYW}) shows that it is required that $Z\gtrsim 3$ to get from the $T$ parameter lower bounds on KK-masses $\mathcal O(1-2)$ TeV. Using now the model parameter values for the metric $A(y)$ in (\ref{A}) and a Higgs profile of the form $h(y)\sim e^{aky}$ we can now evaluate the $Z$ factors from Eq.~(\ref{Z}) in terms of the parameters $(a,\nu,\Delta)$, while we will fix the total warp factor as $A(y_1)=35$. To see the dependence of $Z$ on the various parameters we plot in Fig.~\ref{fig1} the contour levels of $Z$ in the planes ($\Delta$,$\nu$) and ($a$,$\nu$).
%

%%%%%%%%%%%%%%%%%%%%%%%%%%%%%%%%%%%%%%%%%%%%%%%

\section{The Hierarchy Problem}

\label{hierarchy}

As we have found at the end of Sec.~\ref{model} the $Z$ factors can become large for an exponential Higgs profile if the parameter $a$ is ``small". From a holographic dual point of view this can be translated to a small dimension for the Higgs condensate in the IR. This of course raises the question to which extent this reintroduces the hierarchy problem that we claimed to have solved by Higgs compositeness. For instance in the context of RS models this question has been discussed in Ref.~\cite{Luty:2004ye} whose main lines we follow essentially here.
In order to determine whether the hierarchy problem is successfully solved by a profile of the form $h(y)\sim e^{aky}$ with small $a$ let us write the solution of our Higgs profile subject to a generic electroweak symmetry preserving UV-boundary condition $h'(0)=M_0h(0)$ as
\be
h(y)=h_0e^{a k y}\left(1+[M_0/k-a]\int_0^y e^{4A(y')-2a k y'}\right)\,.
\label{fullsoln}
\ee

In the pure AdS case, $A(y)=ky$, the integral yields
\be
h(y)=h_0\left(\frac{M_0/k+a-4}{2(a-2)}e^{a k y}
-\frac{M_0/k-a}{2(a-2)}e^{(4-a) k y}\right)
\label{RSsoln}
\ee
The observation in RS is that for $a>2$ no fine-tuning is necessary in order to keep only the first term, since near the IR brane (where EWSB occurs) the second term is always irrelevant. On the contrary, for $a<2$, the second term would be dominating and one needs to fine-tune $M_0/k=a$ in order to maintain the solution $h(y)\sim e^{aky}$. This fact has a simple holographic interpretation: since $\dim(\mathcal O_H)=a$ the hierarchy problem is solved by compositeness of the Higgs for $a>2$, but not for $a<2$ (see Ref.~\cite{Luty:2004ye} for a more detailed discussion).

In our case the situation is similar. Again for $a<2$ the solution $h(y)\sim e^{a k y}$  will  be fine-tuned due to the exponential enhancement in the integrand. Moreover now even for $a>2$ one has to be careful. Let us rewrite the solution as
\bea
h(y)&=&h_0 e^{aky} \left[1 +(M_0/k-a)
\int_0^y e^{-2(a-2)ky'}\left(1-\frac{y'}{y_s}\right)^{-\frac{4}{\nu^2}}
\right]
\\
\label{eq:hyF}
&=&h_0 e^{aky} \bigg[
1 + (M_0/k -a) \left[ F(y) - F(0) \right]
\bigg]
,
\eea
where
\begin{equation}
F(y) =  e^{-2(a-2) k y_s} y_s \left[ -2(a-2) k y_s \right]^{-1 + 4/\nu^2} \Gamma \left[ 1 - \frac{4}{\nu^2} , -2(a-2) k( y_s - y) \right].
\end{equation}
Note that $F(y)$ is defined as a complex function but its imaginary parts cancels in \eqref{eq:hyF} leading to a real solution. One should view $F(y)$ as the generalization of $F_{\rm RS}(y)=e^{-2(a-2)ky}$ in the RS case.

\begin{figure}[p]
\centering
\begin{minipage}[c]{0.19\textwidth}
~~~~~~~~\textbf{(a)}
\end{minipage}
\begin{minipage}[c]{0.8\textwidth}
 \begin{psfrags}
  \input{delta1-psfrag.tex}
  \includegraphics[width= 0.83\textwidth]{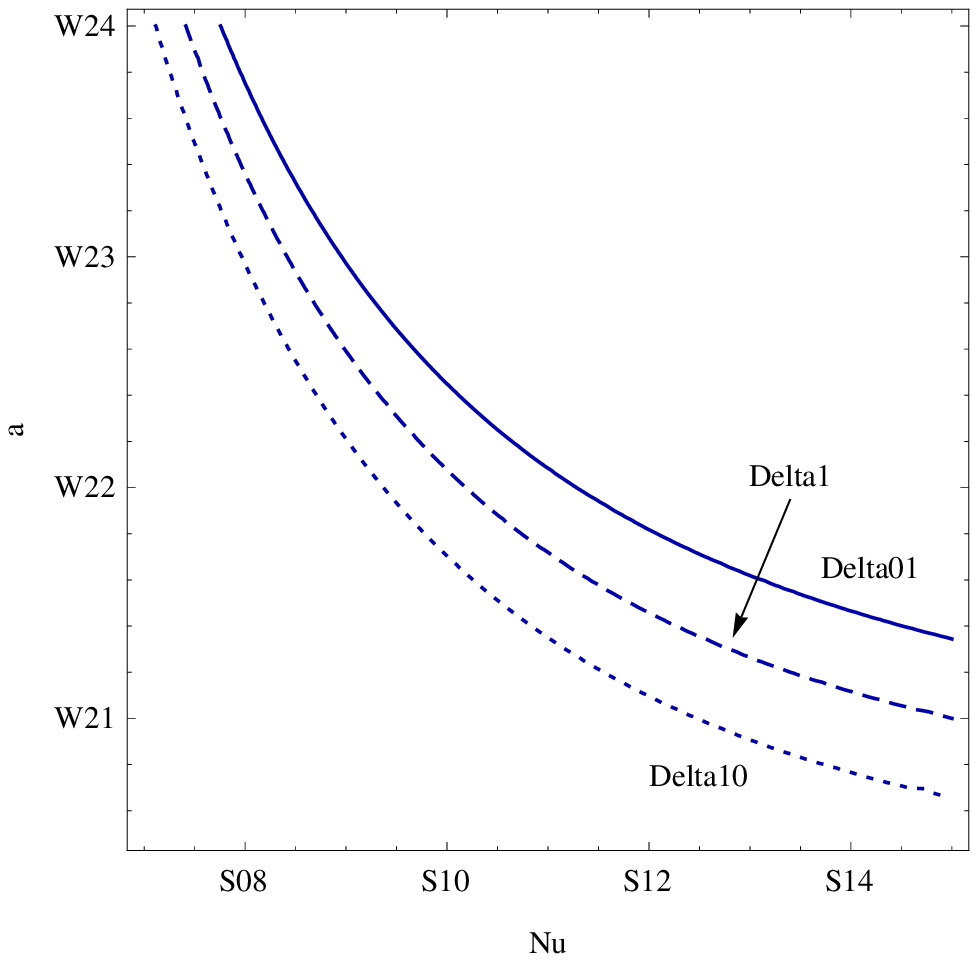}
 \end{psfrags}
\end{minipage}
\\ \vspace{5mm}
\begin{minipage}[c]{0.19\textwidth}
~~~~~~~~\textbf{(b)}
\end{minipage}
\begin{minipage}[c]{0.8\textwidth}
 \begin{psfrags}
  \input{delta2-psfrag.tex}
  \includegraphics[width= 0.83\textwidth]{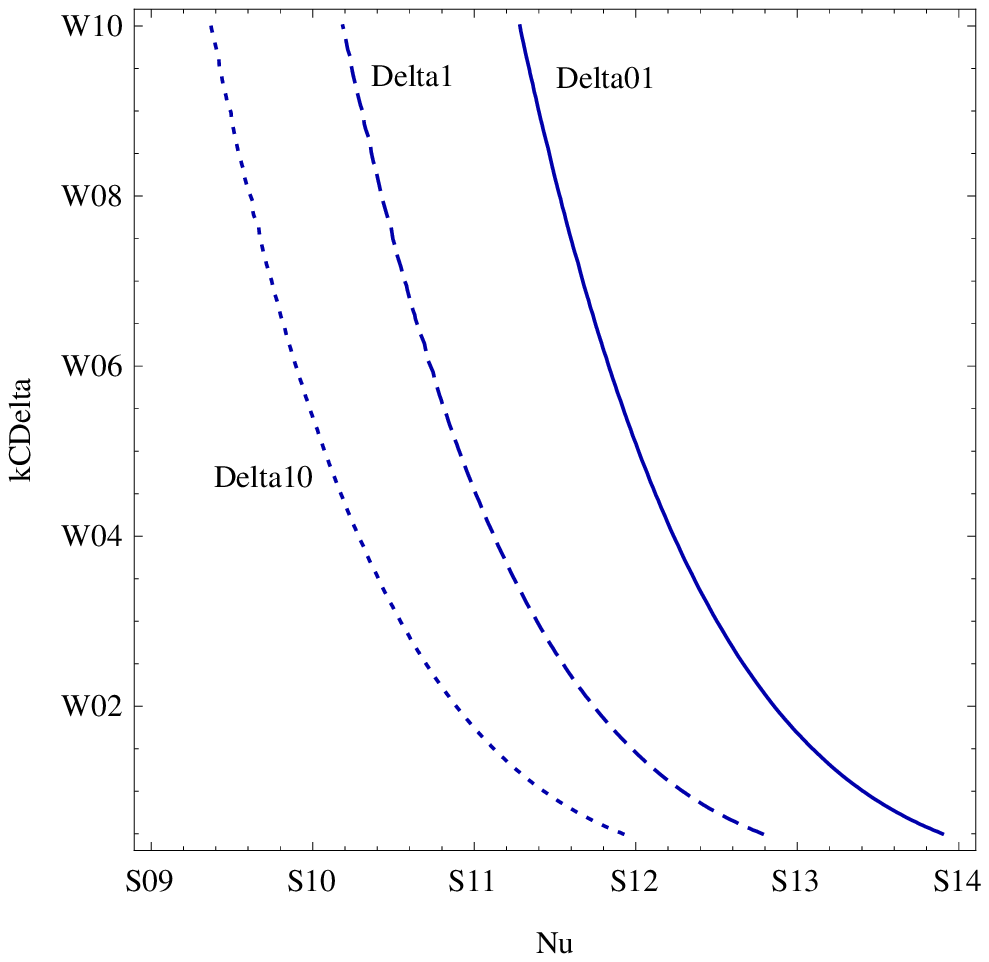}
 \end{psfrags}
\end{minipage}
\caption{\it Contour levels of $\delta$ for $A(y_1) = 35$: \textbf{(a)} as a function of $\nu$ and $a$ for $k\Delta=1$; and \textbf{(b)} as a function of $\nu$ and $k\Delta$ for $a=2.2$.}
\label{fig2}
\end{figure}

Similarly to the AdS case, in order to keep the exponential solution without the need of a fine-tuning we must require the function $F(y)$ to be small. Since $F$ is a monotonically increasing function of $y$ it will be enough to inspect $F(y_1)$. In order to quantify this let us define 
\begin{equation}
\delta \equiv \left\vert F(y_1) \right\vert ,
\label{definitiondelta}
\end{equation}
which will be a measure of the fine-tuning required in $(M_0/k - a)$ in order to keep the exponential solution. In particular the absence of fine-tuning requires roughly $\delta \lesssim \mathcal O(1)$. $\delta$ is a decreasing function of $a$, $\nu$ and $\Delta$, so we need to impose a lower bound on $a=a_0(\nu,\Delta)$ below which one would need to fine-tune $M_0/k\simeq a$ in order to keep the simple exponential solution that improves the EWPT. The behavior of $\delta$ as a function of $a$ and $\nu$ with $k\Delta=1$ is plotted in Fig.~\ref{fig2}.

We can see that the lower bound on $a$ lies (depending on the value of $\nu$) a little above $a=2$, but not by much in the shown parameter range. One can then reinterpret this as stating that for a given $a>2$ there will be a curve $\Delta(\nu)$, as it is shown in Fig.~\ref{fig2}, below which keeping the exponential solution amounts to a fine-tuning. In particular it will be inconsistent to blindly take the limit $\Delta\to 0$.

There is here again a simple holographic interpretation.\footnote{See Refs.~\cite{Vecchi:2010aj} for related ideas.} The dimension of the Higgs condensate corresponding to the solution $h(y)\sim e^{aky}$ depends on $y$. Since the renormalization group (RG) scale is given by the warp factor we have
\be
\dim(\mathcal O_H)=\frac{h'}{h\,A'}=\frac{a}{1+\frac{1}{k(y_s-y)\nu^2}}\,.
\ee
Starting in the UV with some $\dim(\mathcal O_H)>2$ as required to avoid the fine-tuning and solve the hierarchy problem by a composite Higgs, the Higgs mass term $|\mathcal O_H|^2$ will have dimension~\footnote{We use the fact that in the large $N_c$ limit operator products become trivial.} $\dim(|\mathcal O_H|^2)=2\dim(|\mathcal O_H|)>4$  and will be an irrelevant operator becoming more and more suppressed along the RG flow. However following the RG flow further the theory departs from the conformal fixed point, $\dim(\mathcal O_H)$ decreases and there will be a critical RG scale $\mu_c$ at which $\dim(\mathcal O_H)<2$. As a consequence $|\mathcal O_H|^2$ will become a relevant operator and will start increasing again. As long as this happens far enough, near the IR, there is no concern as, at the scale $\mu_c$, the mass term is really small and there is simply not enough RG time for it to become large enough before EWSB occurs. On the other hand a low dimension of the condensate is essential to generate sizable wave function renormalizations for the light Higgs mode that will eventually allow us to suppress the $S$ and $T$ parameters.

%%%%%%%%%%%%%%%%%%%%%%%%%%%%%%%%%%%%%%%%%%%%%%%

\section{Bounds from Precision Observables}
\label{results} 

In this section we will present the bounds on the lightest new states that appear in our model, namely the lightest ($n=1$) KK modes corresponding to the fields that propagate in the bulk. For simplicity we will restrict ourselves to the case in which the fermions are localized on the UV brane. In this case the strongest constraints on the masses of new KK modes in the theory will be given by the dominant $S$ and $T$ parameters, which can be computed from Eq.~\eqref{STYW}. The current experimental bounds on these parameters, for an SM fit with a reference Higgs mass of $117~\mathrm{GeV}$ and assuming $U=0$ are~\cite{Nakamura:2010zzi}
\begin{equation}
T = 0.07 \pm 0.08, 
~~~
 S = 0.03 \pm 0.09 
.
\label{TS}
\end{equation}
with a correlation between $S$ and $T$ of $87\%$ in the fit.

There are four free parameters of our model: $y_1$, $\Delta$, $a$ and $\nu$. In order to better grasp the significance of our results we will trade these parameters for other quantities with a clearer physical meaning~\footnote{The results for this model can be found in terms of these original parameters in Ref.~\cite{Cabrer:2010si}.}. As usual we will fix the Planck--weak hierarchy by setting the warp factor $A(y_1)=35$, which imposes a functional relation $y_1 = y_1 (\Delta,\nu)$ so that $y_1$ increases with $\Delta$ and $\nu$. Moreover we are interested in describing our model in terms of the 5D curvature radius over the interval, $L(y)$, which can be read from Eq.~\eqref{curvature-res} and is plotted in Fig.~\ref{fig0}. Restricting ourselves to the region where $L(y)$ is a monotonic function, as discussed in Sec.~\ref{model}, the minimal curvature radius will be attained at the IR brane. We can then use Eq.~\eqref{curvature-res} to use $L(y_1)\equiv L_1$ as a new parameter, trading it for $\Delta$. Perturbativity in the 5D gravity theory will be under control as long as $M_5L_1\gtrsim 1$ and since $kL_1<1$ for any value of the parameters which departs from AdS~\footnote{The smaller $kL_1$ the larger the deformation of AdS in the IR.} it turns out that $M_5>k$  which may lead to a mild hierarchy between $M_5$ and $k$. In order to avoid this hierarchy to grow too large we shall restrict ourselves to values $kL_1\gtrsim 0.2$~\footnote{We will quantify more precisely this statement at the end of this section.}.
Finally in order to account for the model to solve the hierarchy problem, as discussed in Sec.~\ref{hierarchy}, we will trade $a$ for the fine-tuning parameter $\delta$ defined in Eq.~\eqref{definitiondelta}. The Higgs solution is free of fine-tuning when $\delta \lesssim \mathcal{O}(1)$.

\begin{figure}[t]
\centering
\begin{psfrags}
\input{boundsnu-psfrag.tex}
\includegraphics[width=0.8\textwidth]{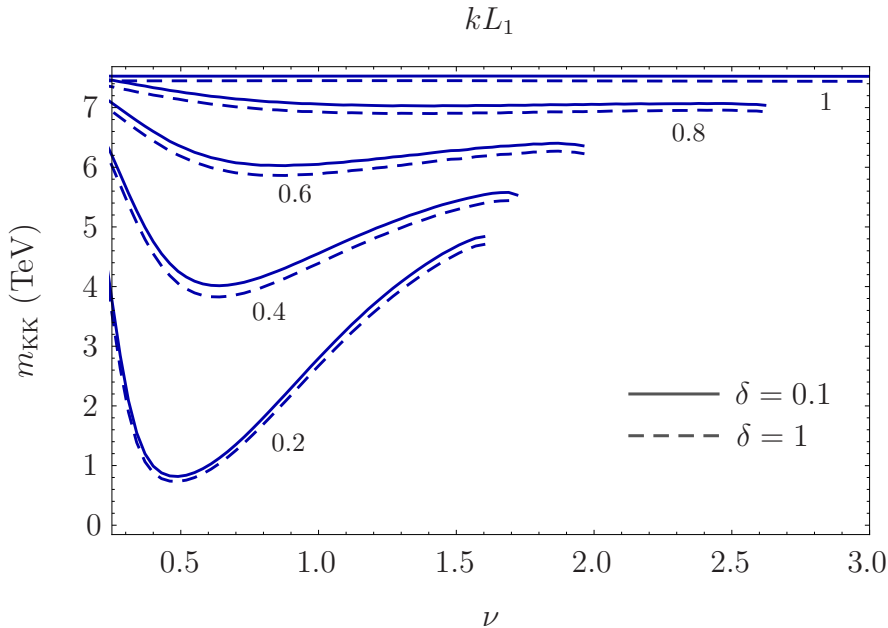}
\end{psfrags}
\caption{\it Plot of $95\%$ CL lower bounds on the first KK-mode mass of gauge bosons as a function of $\nu$ for $A(y_1) = 35$ and different values of the IR curvature radius $L_1$ and the fine-tuning measure $\delta$. The masses corresponding to the three kinds of gauge bosons have a splitting smaller than $1\%$. The region where curves stop corresponds to the excluded region of Fig.~\ref{fig0}. 
}
\label{fig:bounds-Ldeltanu}
\end{figure}

So we are left with the new set of parameters: $kL_1$, $\delta$ and $\nu$ [after fixing $A(y_1)= 35$]. The mass eigenvalue for gauge boson KK modes is given as the solution to Eq.~(\ref{f}) and the corresponding bounds from the experimental value of $T$ and $S$, Eq.~(\ref{TS}), using this set of parameters are shown in Fig.~\ref{fig:bounds-Ldeltanu}. We can see that lowering the minimum curvature radius in the interval of our model yields softer bounds on KK modes. This is an expected result since lower curvatures need smaller values of $\Delta$ (as can be read from Fig.~\ref{fig0}) providing in turn larger values of $Z$. Since we have chosen small enough values of $\delta$, the fact that the model solves the hierarchy problem is guaranteed. Being more demanding with respect to fine-tuning requirements, i.~e.~imposing $\delta$ to be smaller, yields stronger bounds since a smaller value of $\delta$ implies a larger $a$ (Fig.~\ref{fig2}) which in turn makes $Z$ to decrease. However, the effect of changing $\delta$ has only a minor impact on the bounds. Finally the behavior of the curves with $\nu$ is the result of a compromise between the fact that smaller values of $\nu$ correspond to larger values of $Z$ and that, for a constant IR curvature radius $L_1$, lowering $\nu$ implies increasing $\Delta$ as shown in Fig.~\ref{fig0}.

\begin{figure}[h]
%\begin{figure}[p]
\centering
\begin{psfrags}
\input{minbounds-psfrag.tex}
\includegraphics[width=0.8\textwidth]{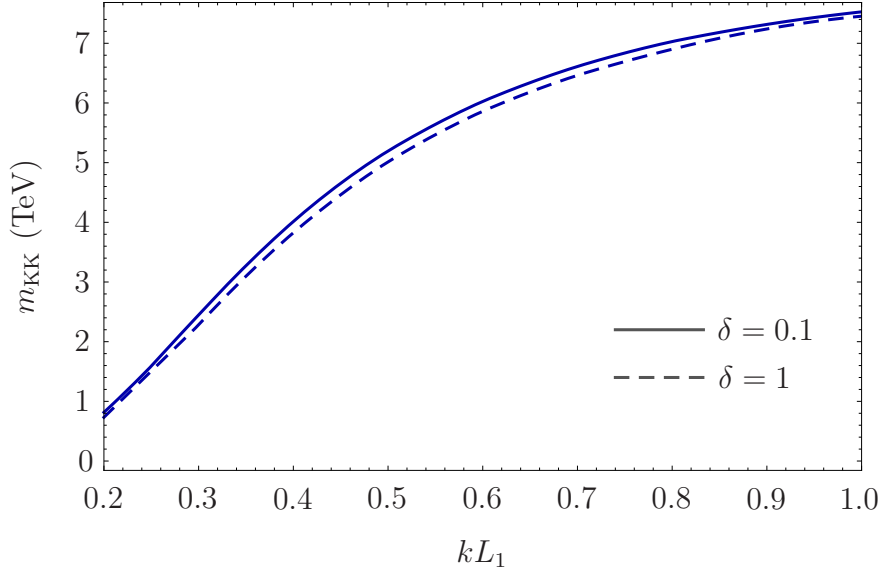}
\end{psfrags}
\caption{\it Plot of the minima in Fig.~\ref{fig:bounds-Ldeltanu} as a function of the IR curvature radius $L_1$. }
\label{fig:bounds-L}
\end{figure}

Since the bounds in Fig.~\ref{fig:bounds-Ldeltanu} show minima as a function of $\nu$ we can eliminate this parameter in order to plot the lower bound on the KK mass as a function of $L_1$ for fixed values of $\delta$. The result is shown in Fig.~\ref{fig:bounds-L}. %

We can see that we can get bounds as low as $m_{KK} \gtrsim 1~\mathrm{TeV}$. The origin of the results shown in Fig.~\ref{fig:bounds-L} is twofold. The smaller $L_1$ the larger the $Z$-factor which in turn triggers:
\begin{itemize}
\item
First a decreasing dependence of $S$ and $T$ with respect to $\rho$ and lower bounds on $m_{KK}$.
\item
Second a smaller angle defined by $\vartheta=\tan(T/S)$ and consequently, using the large correlation between the $S$ and $T$ parameters, a path which approaches the major axis of the $95\%$ CL ellipse allowing for yet lower bounds on $m_{KK}$.
\end{itemize}

\begin{figure}[htb]
%\begin{figure}[p]
\centering
\begin{psfrags}
\input{ellipse-psfrag.tex}
\includegraphics[width=0.85\textwidth]{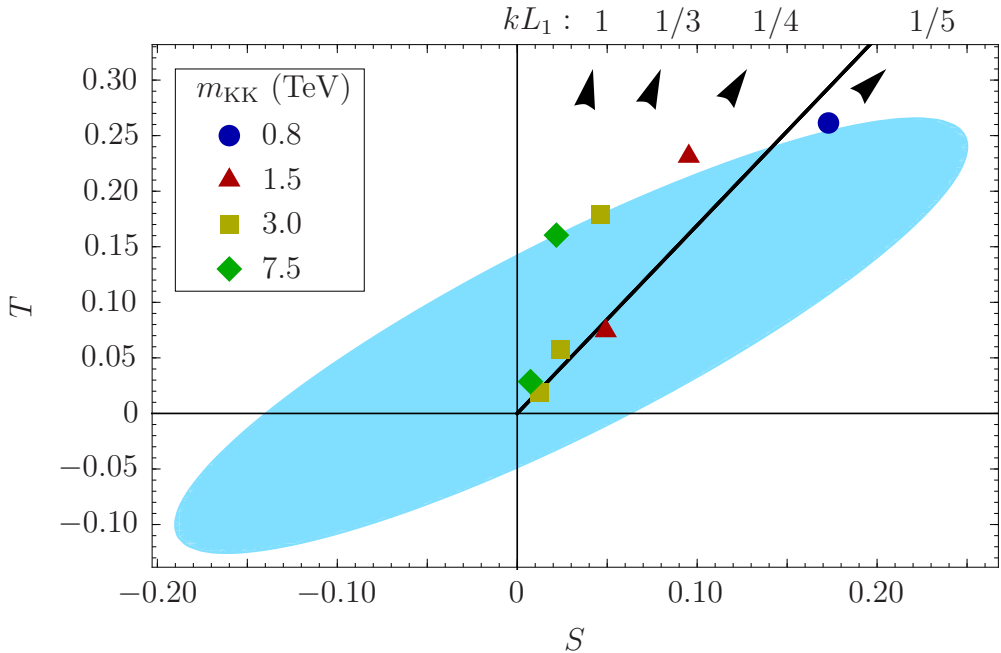}
\end{psfrags}
\caption{\it Plot of values of $T$ and $S$ for different values of $L_1$ (different rays) and  $m_{KK}$ (dots along the rays). The $95\%$ CL ellipse corresponding to 2 d.o.f., where the bounds can be obtained from, is superimposed. The arrows indicate decreasing values of $m_{KK}$ starting from infinity at the origin.}
\label{fig:ellipse}
\end{figure}

\begin{figure}[p]
\centering
\begin{psfrags}
\input{WY-psfrag.tex}
\includegraphics[width=0.8\textwidth]{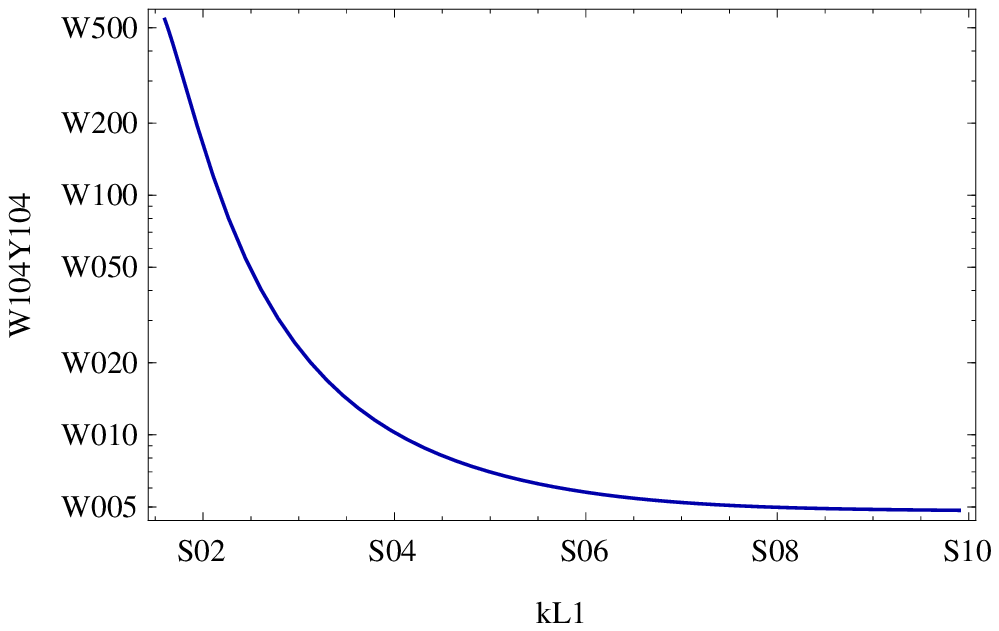}
\end{psfrags}
\caption{\it Plot of the observables $W=Y$, along the minima in Fig.~\ref{fig:bounds-Ldeltanu}, as a function of the IR curvature radius $L_1$. }
\label{fig:WY-L}
\end{figure}

These two effects are exhibited in Fig.~\ref{fig:ellipse} where we show, in the $(T,S)$ plane, the different paths corresponding to different values of $kL_1$ and the corresponding values of $T$ and $S$ for different values of $m_{KK}$. We can see that for values of $kL_1\simeq 1$ the rays go mainly along the $T$ axis while for small values of $kL_1$ the rays get a longer path before getting off the $95\%$ CL ellipse along its major axis. It should be clear from the plot that the second effect is less important, but non-negligible.
At KK masses of $\mathcal O(1)$ TeV one should also be concerned by the fact that the $W$ and $Y$ parameters do not become too large. 
Moreover the lowest bounds also correspond to relatively small values of the volume $ky_1$~\footnote{See e.g.~Tab.~\ref{tablevalues} in Sec.~\ref{conclusion}.}. Since the $Y$ and $W$ observables in Eq.~(\ref{STYW2}) are volume suppressed and correspond to four-fermion effective operators generated by the exchange of KK-gauge bosons, a natural concern should be whether these observables should stay below their experimental bounds. In our case (where $W=Y$) a fit to all observables for a light Higgs yielded~\cite{Barbieri:2004qk} $Y\simeq W\lesssim 10^{-3}$ at 95\% CL. We plot in Fig.~\ref{fig:WY-L} the values of $W=Y$ as a function of $kL_1$. 
\begin{figure}[p]
\centering
\begin{psfrags}
\input{boundsall-psfrag.tex}
\includegraphics[width=0.835\textwidth]{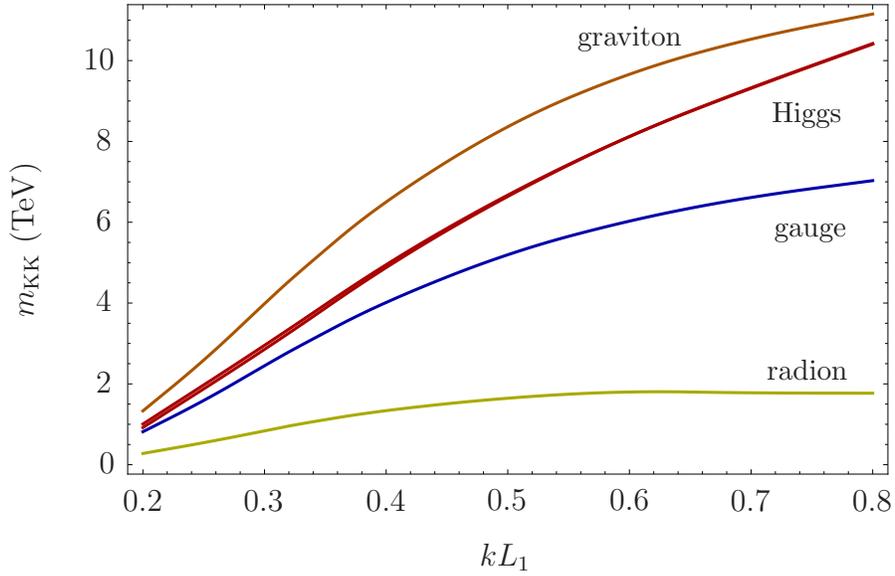}
\end{psfrags}
\caption{\it The same as in Fig.~\ref{fig:bounds-L} for the different fields in the bulk as a function of the IR curvature radius $L_1$. The bounds here correspond to the coordinates of the minima in Fig.~\ref{fig:bounds-Ldeltanu}, irrespectively of each field. The curve for the Higgs includes the pseudoscalar field, the bound of which is around $1\%$ larger for small values of $L_1$. }
\label{fig:bounds-Lfields}
\end{figure}
We see that in the considered range in Fig.~\ref{fig:bounds-L}, $kL_1\gtrsim 0.2$, the values of these observables are well below the experimental bounds. Of course for smaller values of $m_{KK}$ these observables would start to compete with $S$ and $T$ making the bound in Fig.~\ref{fig:bounds-L} to increase at some point for $kL_1 < 0.2$, outside our considered range.

Finally in Fig.~\ref{fig:bounds-Lfields} we also present the bounds on KK masses for the different fields living in the bulk and compare them with the bounds on gauge boson KK modes. In particular we present the first heavy KK mode mass for the physical Higgs and pseudoscalar, Eqs.~(\ref{Higgsbulk}) and (\ref{eq.eta}), for the graviton, Eq.~(\ref{graviton}), and for the radion, Eq.~(\ref{radiongen}). 

\begin{figure}[p]
\centering
\begin{psfrags}
\input{M5k-psfrag.tex}
\includegraphics[width=0.84\textwidth]{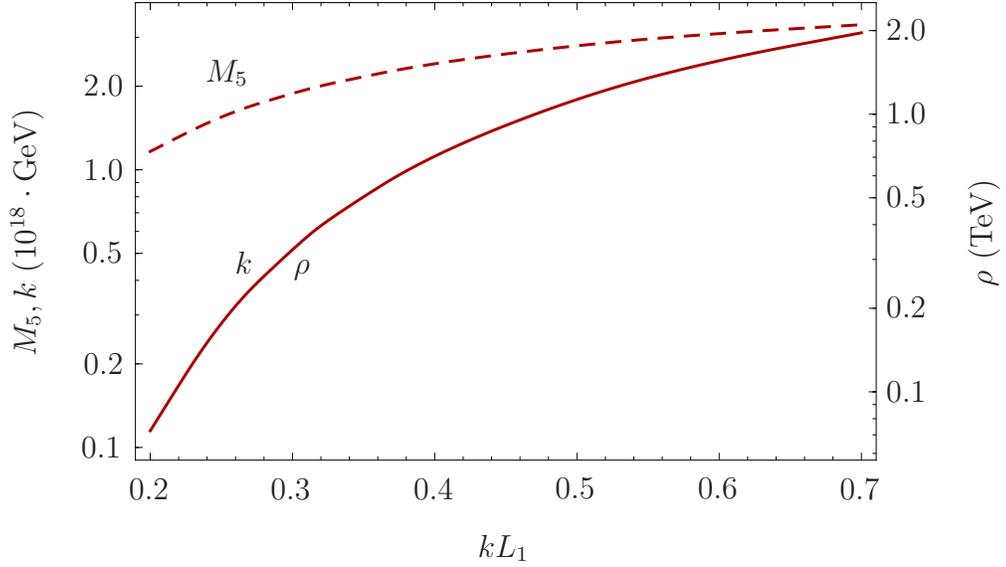}
\end{psfrags}
\caption{\it $M_5$ and $k$ using the bounds on $\rho$ that can be read from Fig.~\ref{fig:bounds-L}, where $\delta = 0.1$ and $A(y_1) = 35$. On the right vertical axis the value of $\rho$ in TeV can be read.}
\label{fig:M5k}
\end{figure}

Let us now make a few comments on the involved scales in our theory. For every set of variables $(\nu, \Delta, a)$ EWPO fix a lower bound on the parameter $\rho$ so that, since we have fixed the total warp factor by $A(y_1)=35$, it turns out that, for every value of $\rho$, $k$ is fixed as $k=e^{35}\rho$.~\footnote{Notice that this procedure is purely operational. We could as well have fixed $k$ (or even the volume $ky_1$) and have considered different warp factors for every case. Physics should not depend on the chosen procedure.} On the other hand the 5D Planck scale can be deduced from Eq.~(\ref{relacion}). Considering the minimal lower bounds on $\rho$ provided by the plot in Fig.~\ref{fig:bounds-Lfields} we obtain the values of $M_5$, $k$ and $\rho$ (which is plotted on the right vertical axis) shown in Fig.~\ref{fig:M5k}.
\begin{figure}[p]
\centering
\begin{psfrags}
\input{higgsft-psfrag.tex}
\includegraphics[width=0.76\textwidth]{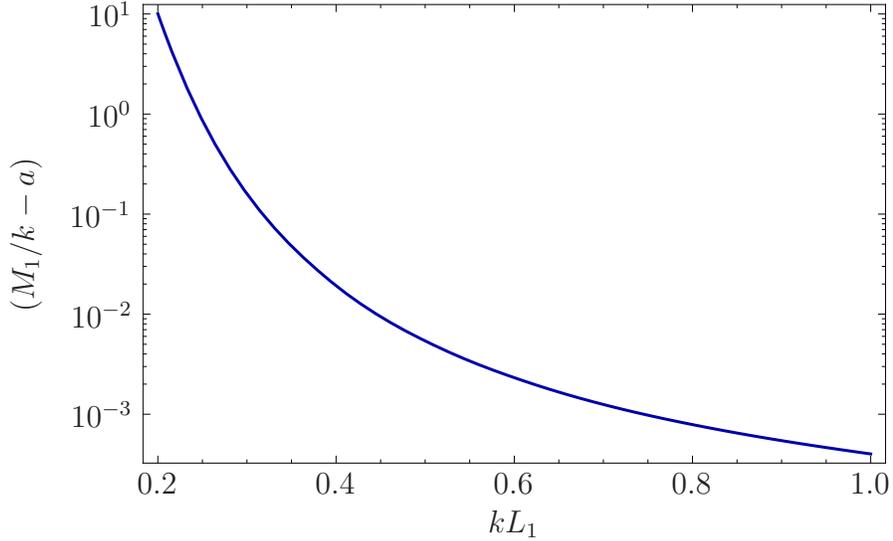}
\end{psfrags}
\vspace{2mm}
\caption{\it $(M_1/k - a)$ for a light Higgs mode with a mass of $120~\mathrm{GeV}$, using the bounds on $\rho$ that can be read from Fig.~\ref{fig:bounds-L}, where $\delta = 0.1$. This quantity can be understood as the amount of fine-tuning required to have a Higgs with this mass.}
\label{fig:higgsft}
\end{figure}
From Fig.~\ref{fig:M5k} we can see that when the IR curvature radius $kL_1$ decreases by a factor of $\sim 5$, $\rho$ (and $k$) decreases by a factor $\sim 20$ while $M_5$ only decreases by a factor $\sim 2$. The parameter $M_5L_1=(M_5/k)kL_1$ which controls perturbativity in the 5D gravity theory thus \emph{increases} by a factor $\sim 2$. This counter-intuitive result comes from the fact that the suppression (enhancement) in the curvature radius (curvature) is overcompensated by the suppression in the value of $k$. This produces a small hierarchy (one order of magnitude) between $k$ and $M_5$. Lowering $kL_1$ further would translate into a larger hierarchy which in turn would translate into a subsequent amout of fine-tuning.

One should keep in mind that the scale $k$ is really a free parameter of the theory that only comes out here as a prediction because we have fixed the volume by the condition $A(y_1)=35$ throughout our analysis. Slightly increasing the volume (and hence $A_1[y_1]$) does not change $S$ and $T$ and we could have used this freedom to fix $k$ such that e.~g.~$M_5/k\sim 5$  and hence $M_5 L_1\sim 1$. 
Let us also mention that the volume is generally reduced in our model since the deformation of AdS is positive, $A(y_1)>ky_1$ (see e.g.~Tab.~\ref{tablevalues}). However it is not the main effect in improving $T$ and moreover the hierarchy is fixed by $A(y_1)$, not $y_1$~\footnote{Of course by relaxing the hierarchy requirement one can easily lower the bounds from EWPT. The simplest examples are the so-called ``little RS models"~\cite{Davoudiasl:2008hx} which solve the hierarchy problem up to scales much below the Planck mass. For instance in these models the requirement $T\sim S$ for an IR localized Higgs (a bulk Higgs with $a\simeq 2$) provides the volume condition $ky_1\sim 4c^2_W$ ($ky_1\sim 9c^2_W$) which translate into mass stabilization till scales of $\sim 30 $ TeV ($\sim 1200$ TeV).}. 

Finally the light Higgs mass is given by Eq.~(\ref{Higgsbulk}) and an approximate analytical expression is given by, see Eq.~(\ref{mH}),
\be
m_H^2
=\frac{2} {Z}\left(M_1/k-a \right)\rho^2
\ .
\label{mHfinal}
\ee
which is written as a function of $\rho$ and the mass parameters $M_1$ and $k$. The Higgs mass is a free parameter in our theory since it is proportional to $\rho^2$ and the prefactor $M_1/k-a$. The only issue here is how much fine-tuning between the tree-level parameters is required in order to have it fixed to a particular value. Clearly if we require $m_H\simeq \rho$ (a heavy Higgs) no fine-tuning is ever necessary. However in that case our theory looks like Higgsless and radiative corrections involving the Higgs would contribute to the EWPO putting the SM clearly outside the experimental bounds. In that case new physics is required to reconcile the theory with EWPT. On the contrary if we require a light Higgs [not too far away from the assumed value for the bounds in Eq.~(\ref{TS})] it will have no effect on the EWPT but it might require some fine-tuning in the tree level parameters. This is the case of the AdS model. However in our model there are two separate effects which favor a light Higgs, as can be seen from Eq.~(\ref{mHfinal}): {\bf i)} the suppresion in the required value of $\rho$ and; {\bf ii)} the enhancement in the value of $Z$. We can quantify this fine tuning by plotting the prefactor $M_1/k-a$ as a function of $kL_1$ using the values of parameters in Fig.~\ref{fig:bounds-Lfields} as we have done in Fig.~\ref{fig:higgsft}.
For values of $M_1/k-a$ of $\mathcal O(1)$ there is no fine-tuning, for values of $\mathcal O(0.1)$ there is a 10\% fine-tuning and so on and so forth. Then we can see that for the region of parameters where we find bounds $m_{KK}\gtrsim 1-2$ TeV there is no fine-tuning while in the RS case the fine-tuning would amount to more than 0.1\%.

%%%%%%%%%%%%%%%%%%%%%%%%%%%%%%%%%%%%%%%%%%%%%%%
\section{Comments on Soft Walls}
\label{sw}

As it has already been mentioned the metric in Eq.~(\ref{A}) has a curvature singularity at a point $y_s>y_1$, i.e.~beyond the IR brane and outside the physical interval. In the absence of an IR brane this singularity becomes a naked one and the associated space is referred to as a soft wall~\cite{Cabrer:2009we,AdS/QCD, Falkowski1,Falkowski2,Falkowski:2009uy, Batell2,Delgado:2009xb,MertAybat:2009mk,Atkins:2010cc,vonGersdorff:2010ht}.
One might wonder what happens if one removes the IR brane and considers a true SW background. A reduction of the $S$ parameter in models with custodial symmetry in a different class of SW background was indeed discovered in Ref.~\cite{Falkowski2} and, from our general considerations, it can be expected that $T$ will be even more reduced than $S$.
Soft walls show some other interesting features such as a greater variety of KK spectra: in particular the density of states above the first KK excitation is typically higher than in two-brane models (and can even be continuous) and could lead to very interesting collider signatures.

Removing the IR brane in our setup requires some care. Our metric can be viewed as a good SW background only for $\nu<2$~\cite{Cabrer:2009we,Gubser:2000nd} and a mass gap only exists for $\nu\geq 1$~\cite{Cabrer:2009we}~\footnote{When $\nu=1$ one encounters a continuous spectrum above the gap, while for $\nu>1$ the spectrum is discrete.}. The parameter space is thus reduced to $1\leq \nu<2$. Moreover up to now we have kept $\rho/k= e^{-A(y_1)}$ fixed in our analysis. This vanishes in the limit $y_1\to y_s$ and one should instead define the Planck-weak hierarchy differently, e.g. by $\rho/k= e^{-ky_s}(ky_s)^{-\frac{1}{\nu^2}}$~\cite{Cabrer:2009we}. 

However without an IR brane and with a bulk Higgs we need to find an alternative location where to trigger EWSB.  The only sensible IR boundary condition for the Higgs profile in the singular background is to demand regularity of the solution. With a linear bulk EOM and linear UV boundary conditions a nontrivial profile can only arise at the price of a fine tuning: satisfying the boundary conditions at the UV brane fixes a certain linear combination of the two bulk solutions (up to an overall normalization). This solution will in general not be regular at $y_s$ so the only solution is the trivial one $h(y)\equiv 0$ and EWSB does not occur. This does not come as a surprise as we have lost the IR brane with its EWSB potential. The only possibility to obtain a nontrivial profile is then a fine-tuning of some parameters in the Lagrangian. One can see how this fine-tuning arises in the smooth limit $y_1\to y_s$ following the logics of Sec.~\ref{hierarchy}. For a fixed value of $a>2$ there will be a finite $\Delta=y_s-y_1>0$ at which the singular solution (initially suppressed by $e^{-2(a-2)ky}$) resurfaces due to the presence of the singular factor in Eq.~(\ref{eq:hyF}). In this case a fine-tuning of the UV mass $M_0$ is needed in order to select the regular solution. This is the original hierarchy problem. Clearly in the exact SW limit, $\Delta=0$, one needs $M_0=a$ precisely and hence an infinite fine-tuning.

The only way out is to trigger EWSB by nonlinear dynamics on the UV brane~\cite{Batell2} or in the bulk. In the former case the VEV of the Higgs field at the UV brane, $h(0)$, is unrelated to the IR scale generated by the warping~\footnote{In fact $h(0)$ has to be fine tuned to a very small number even though this can be made technically natural as long as the UV brane localized Higgs mass term is Planckian.}.
Breaking in the bulk requires to go beyond a simple quadratic potential in the Higgs field and cannot be tackled analytically. Things could be simplified by integrating over the soft wall to create an effective IR brane at $y=y_1< y_s$ along the lines of Ref.~\cite{vonGersdorff:2010ht}. Form factors arise on the IR brane that mimic the characteristic SW spectra, while the bulk potential near the singularity will generate a nontrivial brane potential that could be used to trigger EWSB~\cite{vonGersdorff:2010ht}. It is however not straightforward to engineer a bulk potential $V(\phi,h)$ that can be totally neglected in the reduced bulk $y<y_1$ and whose only effect is the effective IR brane tension.  Although it can be expected that some of the findings of this paper can be translated to the case of a genuine bulk breaking a quantitative statement is hard to be settled in the absence of a precise and calculable model.

%%%%%%%%%%%%%%%%%%%%%%%%%%%%%%%%%%%%%%%%%%%%%%%

\section{Conclusions}
\label{conclusion}

In this paper we have studied generalizations of RS models which allow, following the ideas of Ref.~\cite{Cabrer:2010si}, to avoid the usual paradigm of custodial gauge symmetries.
In the first part of the paper we have derived a series of general results valid for arbitrary metric and scalar backgrounds. These results include the effective theory of a light bulk Higgs (Sec.~\ref{sec:5DSM}), general expressions for electroweak observables $S$, $T$, $Y$ and $W$ (Sec.~\ref{general}) as well as expressions for the mass of the radion (Sec.~\ref{radion}). We have identified a new contribution $Z$ to the wave function renormalization of the Higgs zero mode which can suppress the tree level Higgs mass and plays a major role in reducing the contributions to the $S$ and $T$ parameters. In holographic language the Higgs wave function renormalization $Z$ can be large if the dimension of the Higgs condensate decreases towards the IR, while staying sufficiently large in the UV to solve the hierarchy problem. For that one needs a strong deviation from conformality in the IR, parametrized by a large back reaction of the stabilizing field on the AdS background metric. Motivated by these general considerations we have elaborated on the model proposed in Ref.~\cite{Cabrer:2010si}. The metric possesses a curvature singularity at $y=y_s>y_1$, i.~e.~outside the physical interval $0\leq y\leq y_1$.
Our model can then be described by three input parameters: the quantities $\nu$ and $\Delta=y_s-y_1$ entering our metric, Eq.~(\ref{A}), and the parameter $a$ defined in Eq.~(\ref{M2}) which corresponds to the UV dimension of the Higgs condensate. 
In practice we prefer to trade the quantity $\Delta$ for the 5D curvature radius $L_1$ at the IR brane location. The location of the IR brane, $y_1$, has been fixed by imposing the Planck-EW hierarchy $A(y_1)=35$. The parameter $k$ [or equivalently $\rho=k e^{-A(y_1)}$] is then computed from requiring consistence with EWPT, while the 5D Planck scale $M_5$ is fixed  by the relation Eq.~(\ref{relacion}). Tree-level contributions to the EWPO  are reduced when $a,\ \nu$ and $L_1$ are lowered. 

We have furthermore examined in detail under which circumstances the hierarchy problem could be reintroduced due to the decreasing effective dimension of the Higgs condensate. We find that in 5D language the presence of the singularity beyond the IR brane produces a would-be singular solution to the Higgs profile that needs to be sufficiently suppressed. As our parameter $a$ becomes too small one needs to fine-tune the UV brane mass term for the Higgs field in order to prevent the singular solution to resurface near the IR brane. This can be viewed as a reintroduction of the hierarchy problem: in the 4D dual description the UV dimension is not sufficiently large for the bare Higgs mass operator to stay small all the way to the IR. Insisting that this does not occur produces a bound $a_0(\nu,\Delta)$ which allows us to eliminate the parameter $a$ by $a_0$ and ensures the original solution to the hierarchy problem to be maintained. 

\begin{table}[ht]
\centering
\begin{tabular}{ | c || c | c | c || c | c | c || c | c | c | c | c | }
\hline
$ \displaystyle  \rule[-1.2em]{0pt}{3.2em} kL_1$ & $\nu
$ & $k\Delta$ & $a$ & $Z$ & $ky_1$ & $M_5L_1$&
$\displaystyle\frac{\rho}{\mathrm{TeV}}$
 & $\displaystyle\frac{m^{(\mathrm{KK})}_{\mathrm{gauge}}}{\mathrm{TeV}}$ & $\displaystyle\frac{m^{(\mathrm{KK})}_{\mathrm{Higgs}}}{\mathrm{TeV}}$ &  $\displaystyle\frac{m^{(\mathrm{KK})}_{\mathrm{grav}}}{\mathrm{TeV}}$ & $\displaystyle\frac{m_{\mathrm{rad}}}{\mathrm{TeV}}$
\\ \hline \hline
0.2 & 0.48 & 1.0 & 3.2 & 6.6 & 22 & 2.0 & 0.072 & 0.82 & 0.92 & 1.3 & 0.28 
\\ \hline

0.3 & 0.55 & 1.3 & 2.8 & 2.1 & 25 & 1.1 & 0.32 & 2.4 & 2.9 & 4.0 & 0.84 

\\ \hline

0.4 & 0.64 & 1.6 & 2.5 & 1.2 & 28 & 0.86 & 0.70 & 4.0 & 4.9 & 6.5 & 1.3

\\ \hline

0.5 & 0.73 & 1.7 & 2.4 & 0.86 & 30 & 0.78 & 1.1 & 5.2 & 6.6 & 8.4 & 1.6

\\ \hline

1 & $\infty$ & $\infty$ & 2.1 & 0.47 & 35 & 0.79 & 3.1 & 7.5 & 12 & 12 & 0

\\ \hline 
\end{tabular}
\caption{\it Values of different relevant quantities at some of the points of Fig.~\ref{fig:bounds-Lfields}, where $\delta= \nobreak 0.1$ and $A(y_1)=35$. The stabilization mechanism in our model dissapears when we take the RS limit, leading to vanishing radion mass as expected.
} 
\label{tablevalues}
\end{table}
For each value of $kL_1$ our bounds are thus a function of $\nu$. Some benchmark points along 
Fig.~\ref{fig:bounds-Lfields} can be found in Tab.~\ref{tablevalues} where we display the explicit parameters $\nu,\ \Delta$, $a$ and $y_1$ for reference. 
The EWPO produce a bound on the IR scale $\rho=ke^{-A(y_1)}$ from which the values for the lowest lying KK resonances for the gauge bosons, the Higgs, and the graviton can be calculated. The dominant bounds come from the $T$ parameter for the shown parameter range (cf.~Fig.~\ref{fig:bounds-L}), except for small $kL_1$ where bounds from $S$ and $T$ become comparable. The $W$ and $Y$ parameters are always subdominant. We also produce in the same table the predicted values for the radion mass, which turns out to be rather high due to the strong deviation from AdS. Despite the large IR backreaction on the metric our approximation Eq.~(\ref{mrad}) works to better than 1\% accuracy~\footnote{For $kL_1=0.2$ we find the rather large backreaction $k^{-2}A''(y_1) \approx 4$. The deviation from the exact numerical eigenvalue drops to around $0.1\%$ when the next to leading order, Eq.~(\ref{NLO}), is included.}. The large deviation from AdS also induces another phenomenon evident from Tab.~\ref{tablevalues}: the KK scales $m^{KK}$ and the IR scale $\rho$ become increasingly separated. In the dual theory this is due to the fact that the explicit breaking of conformal symmetry, induced by the relevant operator dual to the GW field $\phi$, happens at scales larger than $\rho$. It is then this higher scale that determines the mass gap of the resonances.
Due to the spurious singularity the 5D curvature radius $L$ decreases along the fifth dimension. It is important to ensure that it does not become too small in order to avoid issues of perturbativity of the 5D gravity theory. Validity of perturbation theory should be ensured if $M_5L_1\gtrsim 1$.
From Tab.~\ref{tablevalues} it is apparent that $M_5L_1$ actually slightly increases when $kL_1$ is lowered. This is due to the rapid decrease in the bound $\rho$ (and hence $k=e^{A(y_1)}\rho$) for small $k L_1$. While this is welcome in that it ensures validity of perturbation theory, we should be careful not to generate a new hierarchy between $k$ and $M_5$ (or, equivalently, between $k$ and $L_1^{-1}$). We have therefore preferred to discard values of $kL_1<0.2$ in our analysis.

There are many topics that we have left to future research. The most obvious generalization of our setup is to allow for fermions to propagate in the bulk rather than to be restricted to the UV brane. Generating fermion mass hierarchies with wave function profiles \cite{Agashe:2004cp} will result in strong bounds on flavour changing neutral currents (FCNC) and CP violation both in the quark and lepton sectors. Nonetheless since our setup bears a certain resemblance to SW models one could expect some improvement along the lines of Ref.~\cite{huberatkins}. We note that we have already established part of the necessary machinery to tackle this problem as we have computed closed expressions for the dimension-six current-current operators generated by tree-level KK exchange in Sec.~\ref{KK}, valid for arbitrary metric and zero mode wave function. Moving the fermions away from the UV brane we expect the $W$ and $Y$ observables and, in genereral, non-oblique corrections to become more important. 
Another topic that deserves closer investigation is the phenomenology of the radion. 
One unequivocal prediction of our model is a heavy radion due to the large deviation of the metric from AdS in the IR. Fortunately the radion wave function is approximated very well by $F(y)=e^{2A(y)}$ due to the excellent accuracy of the leading approximation for the mass, Eq.~(\ref{mrad}). It should therefore be straightforward to work out its couplings to the other light fields and establish its possible signatures at the LHC. 

%%%%%%%%%%%%%%%%%%%%%%%%%%%%%%%%%%%%%%%%%%%%%%%

\section*{Acknowledgments}
We thank M.~P\'erez-Victoria, A.~Pomarol, E.~Pont\'on and J.~Santiago for discussions.
Work supported in part by the Spanish Consolider-Ingenio 2010
Programme CPAN (CSD2007-00042) and by CICYT-FEDER-FPA2008-01430.  The work of JAC is supported by the Spanish Ministry of Education through a FPU grant. The research of GG is supported by the ERC Advanced Grant 226371, the ITN programme PITN- GA-2009-237920 and the IFCPAR CEFIPRA programme 4104-2.   

%%%%%%%%%%%%%%%%%%%%%%%%%%%%%%%%%%%%%%%%%%%%%%%

\section*{Appendix}
\appendix

\section{Gauge Fluctuations}

\label{fluctuationsgauge}

In this appendix we will study the propagation of gauge bosons in the 5D bulk in the presence of the background provided by the Higgs field and the gravitational metric.

\subsection{Warming up with an abelian theory}
\label{warming}
We will first analyze the gauge fixing in a spontaneously broken abelian 5D theory with a Higgs defined by
\be
H(x,y)=\frac{1}{\sqrt{2}}[h(y)+\xi(x,y)]e^{ig_5\chi(x,y)}
\ee
where $h(y)$ is the $y$-dependent Higgs field background, $\xi(x,y)$ the Higgs fluctuation and $\chi(x,y)$ the Goldstone fluctuation. The 5D action for the gauge field $A_M(x,y)$ and the Goldstone boson is given by
\be
S_5=\int d^4x dy\sqrt{-g}\left(-\frac{1}{4} F_{MN}F^{MN}-|D_M H|^2
\right)
\label{accion}
\ee
where $F_{MN}=\partial_MA_N-\partial_NA_M$, $D_MH=\partial_MH-ig_5A_MH$ and $g_5$ is the 5D gauge coupling with mass dimension $-1/2$. The mass dimension of the 5D fields $h$, $\xi$ and $A_M$ is $3/2$ and that of $\chi$ is $1/2$. The action (\ref{accion}) is invariant under 5D gauge transformations
\begin{eqnarray}
A_M(x,y)&\to& A_M(x,y)+\frac{1}{g_5}\partial_M \alpha(x,y)\nonumber\\
\chi(x,y)&\to&\chi(x,y)+\frac{1}{g_5}\alpha(x,y)
\end{eqnarray}

To quadratic order in the fluctuations $A_M$ and $\chi$~\footnote{We need to consider here only the fluctuations of fields $A_M$ and $\chi$ which mix to each other through the mechanism of electroweak breaking. The Higgs fluctuations $\xi$ will decouple from them and are considered in Sec.~\ref{sec:5DSM}.} the action (\ref{accion}) can be written as
\begin{eqnarray}
S_5&=&\int d^4xdy \biggl[-\frac{1}{4}(F_{\mu\nu})^2-\frac{1}{2} \,e^{-2A}(F_{\mu 5})^2
-\frac{1}{2}M_A^2(\partial_\mu\chi- A_\mu)^2  \biggr.\nonumber\\
&&-\frac{1}{2}M_A^2\,e^{-2A}\biggl.(\chi'- A_5)^2 \biggr] ,
\label{accion2}
\end{eqnarray}
where we have defined the $y$ dependent bulk mass 
\be
M_A(y)=g_5 h(y)e^{-A(y)} .
\ee
The bulk EOM's from action (\ref{accion2}) are
\begin{eqnarray}
\Box A_\mu+(e^{-2A}A'_\mu)'-M_A^2 A_\mu
+\partial_\mu\left\{ M_A^2\chi-(\partial^\nu A_\nu)-(e^{-2A}A_5)' \right\}&=&0\label{eq.A}\\
\Box A_5-\partial^\nu A'_\nu+M_A^2(\chi'-A_5)&=&0\label{eq.A5}\\
\Box \chi- \partial^\nu A_\nu+M_A^{-2}\left\{\left(M_A^2 e^{-2A}\right)(\chi'- A_5)
\right\}'&=&0\label{eq.chi}
\end{eqnarray}
and the boundary conditions are
\begin{eqnarray}
%e^{-2A}
 \left( \partial_\mu A_5 - A'_\mu \right) \vert_{y=0,y_1} = 0 \\
%e^{-4A} h' \vert_{y=0,y_1} = 0 \\
 %e^{-4A} 
 \left( \chi' -  A_5 \right) \vert_{y=0,y_1} = 0   
\label{b.c.}
\end{eqnarray}

We can gauge away the last term in Eq.~(\ref{eq.A}) by the gauge condition
\be
\partial^\mu A_\mu-M_A^2\chi+(e^{-2 A}A_5)'=0.
\label{gaugecond}
\ee
By making the ansatz (the dot product denotes an expansion in modes)
\be
A_\mu(x,y)=\frac{a_\mu(x)\cdot f(y)}{\sqrt{y_1}}
\label{ansatzA}
\ee
the EOM (\ref{eq.A}) becomes
\be
m_f^2 f+(e^{-2A}f')'-M_A^2 f=0
\label{eq.f}
\ee
where the functions $f(y)$ are normalized as
\be
\frac{1}{y_1}\int_0^{y_1}f^2(y)dy=1
\ee
and satisfy the boundary conditions
\be
%e^{-2A}
\left. f'\right|_{y=0,y_1}=0.
\label{eq:boundaryf}
\ee

It is easy to see that the gauge condition remains invariant under the whole set of 5D gauge transformations
\be
\alpha(x,y)=\alpha(x)\cdot f(y)
\ee
where $\alpha_n(x)$ are arbitrary 4D gauge transformations which are the remaining 4D invariances. A quick glance at the action, Eq.~(\ref{accion2}), shows that the Goldstone boson degree of freedom (which couples to $\partial^\mu A_\mu$ in the action) should be defined as
\be
G(x,y)=M_A^2 \chi-\left(e^{-2A} A_5\right)'
\ee
while the remaining degree of freedom is defined as~\footnote{The pseudoscalar modes $K_n$ are physical (they are in particular gauge invariant) and could play an important role in experimentally identifying the Higgs as a bulk field. Their equations of motion have been derived previously in Ref.~\cite{Falkowski1}.}
\be
K(x,y)=\chi'-A_5
\ee
which satisfy the decoupled EOM's [from Eqs.~(\ref{eq.A5}) and (\ref{eq.chi})]
\begin{eqnarray}
\Box G+(e^{-2 A}G')'-M_A^2 G&=&0\label{eq.G}\\
 \Box K+\left[M_A^{-2}\left(e^{-2A}M_A^2K \right)'  \right]'
 -M_A^2K&=&0\ . \label{eq.K}
\end{eqnarray}
Eqs.~(\ref{eq.G}) and (\ref{eq.K}) are satisfied by
\begin{eqnarray}
G(x,y)&=&\frac{m_f\,G(x)\cdot f(y)}{\sqrt{y_1}}\label{ansatzG}\\
K(x,y)&=&\frac{K(x) \cdot\eta(y)}{\sqrt{y_1}}\label{ansatzK}
\end{eqnarray}
where $f(y)$ satisfies Eq.~(\ref{eq.f}), and $\eta(y)$ satisfies the bulk EOM
\be
 m_\eta^2\eta+\left[M_A^{-2}\left(e^{-2A}M_A^2\eta \right)'  \right]'
 -M_A^2\eta=0
\label{eq.eta}
\ee
and the (Dirichlet) boundary conditions
\be
\left.%e^{-4A}
\eta\right|_{y=0,y_1}=0\ .
\label{b.c.D}
\ee
The normalization for $\eta$ will be fixed below.
Notice that in the limit $M_A\to 0$ there is no massless mode since the zero mode would have the (trivial) wave function, consistent with the boundary conditions, $\eta(y)\equiv 0$. Only massive KK modes do appear.

In the 4D theory the degrees of freedom are the gauge field $a_\mu(x)$ the Goldstone boson $G(x)$ and the gauge invariant scalar $K(x)$. They transform under the 4D gauge transformation $\alpha(x)$ as
\begin{eqnarray}
\delta_\alpha a_\mu(x)&=&\frac{1}{g}\partial_\mu\alpha(x)\nonumber\\
 \delta_\alpha G(x)&=&\frac{m_f}{g}\,\alpha(x)\nonumber\\
\delta_\alpha K(x)&=&0
\end{eqnarray}
where the 4D gauge coupling is defined as $g=g_5/\sqrt{y_1}$.
It is easy to obtain the 4D effective Lagrangian upon integration of the $y$-coordinate in the action (\ref{accion2}) by using the decomposition 
\begin{eqnarray}
\sqrt{y_1}\,A_5(x,y)&=&\frac{1}{m_f}G(x)\cdot f'(y)-\frac{M_A^2}{m_\eta^2}\,K(x)\cdot \eta(y)\nonumber\\
\sqrt{y_1}\,\chi(x,y)&=&\frac{1}{m_f}G(x)\cdot f(y)-\frac{1}{m_\eta^2}M_A^{-2}\left(M_A^2e^{-2A}\eta(y)\right)'\cdot K(x).
\end{eqnarray}
In fact after integration over the $y$-coordinate and using the EOM (\ref{eq.f}) and (\ref{eq.eta}) one can write down the 4D Lagrangian as
\be
\mathcal L_{4D}=-\frac{1}{4}(\partial_\mu a_\nu-\partial_\nu a_\mu)^2-\frac{1}{2}(m_f a_\mu-\partial_\mu G)^2 
-\frac{1}{2}(\partial_\mu K)^2-\frac{1}{2}m_\eta^2 K^2
\label{modelag}
\ee
where we have fixed the normalization for the wave function $\eta$ as
\be
\frac{1}{y_1}\int_0^{y_1} M_A^2e^{-2A}\eta^2\,dy=m_\eta^2\,.
\ee
In Eq.~(\ref{modelag}) all the squares are to be understood as summations over modes.

%%%%%%%%%%%%
%\begin{comment}
Notice that although the EOM (\ref{eq.G}) and (\ref{eq.K}) are decoupled they arise from the coupled set (\ref{eq.A5}) and (\ref{eq.chi}) and as such the mass eigenvalues are common. Of course this does not mean that Eqs.~(\ref{eq.f}) and (\ref{eq.eta}) should have the same mass eigenvalues and the puzzle can be resolved by noticing that Eqs.~(\ref{eq.f}) and (\ref{eq.eta}) always admit the trivial solutions $f(y)\equiv 0$ and/or $\eta(y)\equiv 0$. In fact a solution $m_1^2$ and $f^{(1)}(y)$ from Eq.~(\ref{eq.f}) corresponds to the mass eigenfunctions $(f^{(1)}(y),\eta^{(1)}(y)\equiv 0)$ and the corresponding solution $m_2^2$ and $\eta^{(2)}(y)$ from Eq.~(\ref{eq.eta}) corresponds to the mass eigenfunctions $(f^{(2)}(y)\equiv 0,\eta^{(2)}(y))$. Then the effective Lagrangian corresponding to all the modes can be written as
\begin{eqnarray}
\mathcal L_{4D}&=&-\sum_{n_1}\left(\frac{1}{4}(\partial_\mu a_\nu^{(n_1)}(x)-\partial_\nu a_\mu^{(n_1)}(x))^2+\frac{1}{2}m^2_{n_1}(a_\mu^{(n_1)}(x))^2\right.\nonumber\\
&+&\left.\frac{1}{2}(\partial_\mu G^{(n_1)}(x))^2+m_{n_1}(\partial^\mu a_\mu^{(n_1)}(x)) G^{(n_1)}(x)\right)\nonumber\\
&-&\sum_{n_2}\left(\frac{1}{2}(\partial_\mu K^{(n_2)}(x))^2+\frac{1}{2}m^2_{n_2} (K^{(n_2)}(x))^2\right)
\end{eqnarray}
%\end{comment}
%%%%%%%%%%%%%

\subsection{The Standard Model}
\label{theSM}
The generalization to non-abelian gauge theories is straightforward. In particular in the $SU(2)\times U(1)$ Standard Model the gauge and Higgs bosons are introduced in the usual way with a 5D action given by
\be
S_5=\int d^4x dy\sqrt{-g}\left(-\frac{1}{4} (F^i_{MN})^2-\frac{1}{4}(F_{MN}^Y)^2-|D_M H|^2
-V(\Phi,H)
\right)
\ee
where the 5D Higgs field is written as
\be
H=\frac{1}{\sqrt 2}e^{i g_5 \chi} \left(\begin{array}{c}0\\h+\xi\end{array}\right)
\ee
and where the matrix $\chi$ only includes the coset and $g_5$ is the 5D $SU(2)_W$ coupling. The $\xi$ field will again decouple so we consider it separately.
Following the standard notation we have
\be
D_M=\partial_M-ig_5A_M\,,\qquad A_M=\left(
\begin{array}{cc}
s_wA^{em}_M+\frac{c_w^2-s_w^2}{2 c_w}Z_M&\frac{1}{\sqrt 2}W^+_M\\
\frac{1}{\sqrt 2}W_M^-&-\frac{1}{2c_w}Z_M
\end{array}
\right)
\ee
and analogously
\be
\chi=\left(
\begin{array}{cc}
\frac{c_w^2-s_w^2}{2 c_w}\chi_Z&\frac{1}{\sqrt 2}\chi_+\\
\frac{1}{\sqrt 2}\chi_-&-\frac{1}{2c_w} \chi_Z
\end{array}
\right)
\label{chimatrix}
\ee
where the weak angle is defined as in the 4D theory, $t_w\equiv g'_5/g_5=g'/g$.
Expanding the Lagrangian to second order we obtain a straightforward generalization of the abelian case, Eq.~(\ref{accion2})
\be
\mathcal L=\mathcal L^{\gamma}+\mathcal L^{Z}+\mathcal L^W
\ee
with
\begin{eqnarray}
\mathcal L^{\gamma}&=&-\frac{1}{4}(F^{\gamma}_{\mu\nu})^2-\frac{1}{2}e^{-2A}(F^{\gamma}_{\mu 5})^2\\
\mathcal L^{Z}&=&-\frac{1}{4}(F_{\mu\nu}^Z)^2-\frac{1}{2}e^{-2A}(F_{\mu5}^Z)^2-\frac{1}{2}M_Z^2(\partial_\mu\chi_Z-A_\mu^Z)^2\nonumber\\
&&-\frac{1}{2}e^{-2A}m_Z^2(\chi_Z'-A_5^Z)^2\\
\mathcal L^W&=&-\frac{1}{2}F_{\mu\nu}^+F_{\mu\nu}^--\frac{1}{2}e^{-2A}F_{\mu5}^+F_{\mu5}^--M_W^2(\partial_\mu\chi_+-A_\mu^+)(\partial_\mu\chi_--A_\mu^-)\nonumber\\
&&-e^{-2A}M_W^2(\chi_+'-A_5^+)(\chi_-'-A_5^-)
\end{eqnarray}
Here we have defined the 5D $y$-dependent gauge boson masses
\be
M_W(y)=\frac{g_5}{2} h(y)e^{-A(y)}\,,\qquad M_Z(y)=\frac{1}{c_w} M_W(y)\,,\qquad M_\gamma(y)\equiv 0
\label{masas2}
\ee

Now we should proceed as in the abelian case and define the mode expansion for the different gauge bosons $A_\mu(x,y)$ with profiles $f_A(y)$ ($A=W,Z,\gamma$) as in Eq.~(\ref{ansatzA}) and the corresponding pseudoscalars $K_A(x,y)$ with profiles $\eta_A(y)$ as in Eq.~(\ref{ansatzK}) which satisfy [Eqs.~(\ref{eq.f}) and (\ref{eq.eta})]
\begin{eqnarray}
m_{f_A}^2 f_A+(e^{-2A}f'_A)'-M_A^2 f_A&=&0 \label{eqfinalf}\\
 m_{\eta_A}^2\eta_A+\left[M_A^{-2}\left(e^{-2A}M_A^2\eta_A \right)'  \right]'
 -M_A^2\eta_A&=&0 
 \label{eqfinaleta}
\end{eqnarray}
where $M_A$ is defined in Eq.~(\ref{masas2}) and $m_{f_A}$ and $m_{\eta_A}$ the mass eigenvalues which are identified with the physical gauge boson masses. 

%\subsection{The zero mode}
%\label{gaugezero}

In case the lightest mode after electroweak breaking is separated by a gap from the KK spectrum the expansion in powers of $m_{A,0}^2$ defined in Eq.~(\ref{mA0}) can be carried out analogously for its eigenvalue and wave function. 
Making the ansatz
\be
f^0_A(y)=1-\delta_A+\delta f_A(y)
\label{fAequal}
\ee 
and
\be
m_A^2=m_{A,0}^2-\delta m_{A}^2
\ee
One finds
\bea
\delta f_A(y)&=&m_{A,0}^2 y_1\int^y_0 e^{2A}\left(\Omega-\frac{y'}{y_1}\right)\nn\\
\delta_A&=&m_{A,0}^2y_1\int^{y_1}_0 e^{2A}\left(\Omega-\frac{y'}{y_1}\right)\left(1-\frac{y'}{y_1}\right)\nn\\
\delta m_{A}^2&=&
m_{A,0}^4 y_1 \int_0^{y_1} e^{2A}\left(\Omega-\frac{y'}{y_1}\right)^2
\eea
where the function $\Omega$ was defined in Sec.~\ref{general}.

\section{Gauge Boson Propagators}
\label{app:propagators}

In this appendix we calculate the gauge boson propagators at zero momentum. In other words we would like to compute
\be
G_{B_0B_1}(y,y')=\sum_{n\geq 1} \frac{f^n(y)\, f^n(y')}{m_n^2}\,,
\label{defG}
\ee
where $B_i=D,N$ denote Dirichlet or Neumann BC at the boundaries at $y=y_i$:
\be
G_{DB_1}(0,y')=0\quad {\rm or}\quad G'_{NB_1}(0,y')=0\,,
\label{B0}
\ee
and
\be
G_{B_0D}(y_1,y')=0\quad {\rm or}\quad G'_{B_0N}(y_1,y')=0\,,
\label{B1}
\ee
respectively.
The sum excludes any zero mode (if present). The $f_n$ are the wave functions in the symmetric phase
\be
(e^{-2A}f_n')'+m_n^2f_n=0\,.
\label{eomf}
\ee
The propagators then satisfy the EOM
\be
\partial_y\left[e^{-2A(y)}\partial_y G_{B_0B_1}(y,y')\right]=-y_1\delta(y-y')\,,
\label{eomG}
\ee
for $B_0B_1\neq NN$ and
\be
\partial_y\left[e^{-2A(y)}\partial_y G_{NN}(y,y')\right]=1-y_1\delta(y-y')\,,
\label{eomGNN}
\ee
in the case of Neumann-Neumann BC with zero mode subtracted.
These equations are easily derived from Eq.~(\ref{eomf}) using the completeness relation~\footnote{Furthermore recall that our normalization reads $\int_0^{y_1}f_n^2=y_1$, in particular $f_0(y)=1$ in the $NN$ case.}
\be
\sum_{n\geq 0}f_n(y)f_n(y')=y_1\delta(y-y')\,.
\ee
The boundary conditions have to be supplemented by the jump and continuity relations
\bea
e^{-2A(y')}\partial_y \left[G_{B_0B_1}(y'+\epsilon,y')-G_{B_0B_1}(y'-\epsilon,y')\right]&=&-y_1\,,\nn\\
G_{B_0B_1}(y'+\epsilon,y')-G_{B_0B_1}(y'-\epsilon,y')&=&0\,.
\label{jump}
\eea
For $B_0B_1\neq NN$ the solutions are straightforward and read
\bea
G_{DN}(y,y')&=&y_1\int_0^{y_<}e^{2A}\,,\nn\\
G_{ND}(y,y')&=&y_1\int_{y_>}^{y_1}e^{2A}\,,\nn\\
G_{DD}(y,y')&=&y_1\frac{\int_0^{y_<}e^{2A}\cdot\int_{y_>}^{y_1}e^{2A}}{\int_0^{y_1}e^{2A}}\,,
\label{GBB}
\eea
where $y_<$ ($y_>$) denotes the smaller (larger) of the pair $y,y'$. One can immediately verify that these are solutions to the system of Eqs.~(\ref{eomG}), (\ref{B0}), (\ref{B1}) and (\ref{jump}). 
The $NN$ case requires more care. One can always shift the solution by a $y'$ dependent constant: the bulk Eq.~(\ref{eomGNN}) is invariant under such a shift and so are the BC and the conditions Eq.~(\ref{jump}). After imposing symmetry in the interchange of $y$ and $y'$ (obvious from the definition Eq.~(\ref{defG})), one still has an undetermined $y'$ independent constant. In fact one can immediately verify that
\be
G_{NN}(y,y')=\int_0^{y_<}d\hat y\,e^{2A(\hat y)}\hat y
+\int_{y_>}^{y_1}d\hat y\, e^{2A(\hat y)}(y_1-\hat y)+c
\ee
is a solution (symmetric under interchange of $y,y'$) to the system for arbitrary
constant $c$. To fix $c$, we impose that $G_{NN}(0,0)$ reduces to the brane to brane propagator computed in Eq.~(\ref{prop2}):
\be
G_{NN}(0,0)=-\lim_{p\to 0}\left(\frac{1}{\Pi(p^2)}-\frac{1}{p^2}\right)\,.
\ee
Notice in Eq.~(\ref{prop2}) one can set $m_A$ to zero in the symmetric phase. One finds
\be
G_{NN}(0,0)=y_1\int_0^{y_1}d\hat{y}\,e^{2A(\hat y)}\left(1-\frac{\hat{y}}{y_1}\right)^2
\ee
which fixes $c$ uniquely and one ends up with
\be
G_{NN}(y,y')=\int_0^{y_<}d\hat y\,e^{2A(\hat y)}\hat y
+\int_{y_>}^{y_1}d\hat y\, e^{2A(\hat y)}(y_1-\hat y)
-\frac{1}{y_1}\int_0^{y_1}e^{2A(\hat y)}\hat y(y_1-\hat y)\,.
\label{GNN}
\ee
%

%%%%%%%%%%%%%%%%%%%%%%%%%%%%%%%%%%%%%%%%%%%%%%%

\end{document}